\documentclass[12pt,reqno]{article}
\usepackage{amsmath}
\usepackage{amsthm}
\usepackage{amsfonts}
\usepackage{amssymb}
\usepackage{epic}
\usepackage{eepic}
\usepackage{graphics}
\newcommand{\qps}[2]{\big( #1;#2\big) _{\infty }}
\newcommand{\fqps}[3]{\big( #1;#2 \big)_{#3}}
\newcommand{\Li}{\text{Li}_{2}}
\renewcommand{\Re}{\text{Re}\, }
\renewcommand{\Im}{\text{Im}\, }
\newcommand{\Y}{\Upsilon}

\newcommand{\bz}{\bar{z}}
\def\S{{\cal S}}
\def\ts{\tilde s} 
\def\s{\sigma}
\def\ta{\tilde\a}

\def\QR{\mathbb{R}} 
 
\def\QZ{\mathbb{Z}}
\def\Ga{\Gamma} 
\def\a{\alpha}
\def\b{\beta}

\def\e{\epsilon} 
 
\def\R{{\cal R}}
\def\ren{{\rm ren}}

\newcommand{\beq}{\begin{equation}}
\newcommand{\eeq}{\end{equation}}

\def\beas{\begin{eqnarray*}}
\def\eeas{\end{eqnarray*}}
\def\bea{\begin{eqnarray}}
\def\eea{\end{eqnarray}}
\def\a{\alpha} 
\def\b{\beta} 
 
\newcommand{\remlst}{\begin{list}
{(\arabic{num})}{\usecounter{num}\topsep0cm \itemsep0cm \parsep0cm}}
\input epsf
\newcount\figno
\def\fig#1#2#3{
\par\begingroup\parindent=0pt\leftskip=1cm\rightskip=1cm\parindent=0pt
\baselineskip=15pt
\global\advance\figno by 1
\epsfxsize=#3
\centerline{\epsfbox{#2}}
\vskip 12pt
{\bf \small Figure \the\figno:} {\small #1}\par
\endgroup\par
}
\def\figlabel#1{\xdef#1{\the\figno 
\mbox{ }}}
\def\encadremath#1{\vbox{\hrule\hbox{\vrule\kern8pt\vbox{\kern8pt
\hbox{$\displaystyle #1$}\kern8pt}
\kern8pt\vrule}\hrule}}

\title{\bf Boundary Liouville theory at $c=1$}  
\author{\\[5mm] Stefan Fredenhagen \\[5mm] 
Institut des Hautes Etudes Scientifiques,\\
F-91440 Bures-sur-Yvette, France\\[8mm]  
Volker Schomerus \\[5mm]Service de Physique 
Th{\'e}orique, CEA Saclay,\\ F-91191 Gif-sur-Yvette, France\\[5mm]}
\date{September, 2004} 
\begin{document}
\begin{titlepage}      \maketitle       \thispagestyle{empty}

\vskip1cm
\begin{abstract} 
The $c=1$ Liouville theory has received some attention recently 
as the Euclidean version of an exact rolling tachyon background. 
In an earlier paper it was shown that the bulk theory can be 
identified with the interacting $c=1$ limit of unitary minimal 
models. Here we extend the analysis of the $c=1$-limit to the 
boundary problem. Most importantly, we show that the FZZT branes 
of Liouville theory give rise to a new 1-parameter family of 
boundary theories at $c=1$. These models share many features 
with the boundary Sine-Gordon theory, in particular they 
possess an open string spectrum with band-gaps of finite 
width. We propose explicit formulas for the boundary 2-point 
function and for the bulk-boundary operator product expansion 
in the $c=1$ boundary Liouville model. As a by-product of our 
analysis we also provide a nice geometric interpretation for 
ZZ branes and their relation with FZZT branes in the $c=1$ 
theory. 
\end{abstract} 

\vspace*{-21.4cm}\noindent
{\tt {SPhT-T04/121}}\hfill
{\tt {hep-th/0409256}}\\
{\tt {IHES/P/04/42}}
\bigskip\vfill
\noindent
\phantom{wwwx}{\small e-mail:}{\small\tt
vschomer@spht.saclay.cea.fr \ \& \ stefan@ihes.fr} 

\end{titlepage} 

\baselineskip=19pt 
\setcounter{equation}{0} 
\section{Introduction}
\def\tr{{\rm tr}}

The study of tachyons and their condensation processes has been 
a longstanding challenge in string theory. During recent years, 
the interest in branes has added significantly to the relevance 
of this issue since open string tachyons occur in various  
configurations of stable branes, the most important being the 
brane-antibrane system. Fortunately, 
unstable brane configurations turned out to be much more 
tractable than bulk backgrounds with instabilities such as e.g.\ 
the 26-dimensional bosonic string. In fact, various different 
approaches have been employed over the last five years, 
including string field theory (see e.g.\ \cite{Sen:1999nx} and 
\cite{Taylor:2003gn} for a review), effective field theory 
models (see \cite{Garousi:2000tr,Sen:2002in} for some early 
work),  and world-sheet conformal field theory (see e.g.\ 
\cite{Harvey:2000na,Schomerus:2002dc} and references therein). 
\smallskip 

In \cite{Sen:2002nu}, Sen initiated the study of {\em exact 
time-dependent solutions}. One goal of his work was to find a 
world-sheet description of so-called S-branes \cite{Gutperle:2002ai}, 
i.e.\ of branes that are localized in time. In this context he 
proposed to add a $\cosh$-shaped boundary interaction to 
the usual world-sheet model of D-branes in a flat 
space-time,  
\begin{equation} \label{senact}
 S_S[X] \ = \ \frac{1}{4\pi} \int_\Sigma 
                 d^2 z \, \eta_{\mu\nu} \partial X^\mu
                 \bar \partial X^\nu + \int_{\partial \Sigma} 
                 d u \, \lambda \cosh X^0(u) \ \ . 
\end{equation} 
Here, $X^0$ denotes the time-coordinate and we have chosen 
units in which $\a' =1$. With such an interaction term the open 
string tachyon becomes large at early and late times and hence, 
according to an earlier proposal of Sen \cite{Sen:1999xm}, is 
believed to dissolve the brane for $|x_0| \rightarrow \infty$. 
Unfortunately, the world-sheet model (\ref{senact}) appears to be 
ill-defined without an additional prescription to select the proper 
vacuum. To resolve this issue, Sen argued that the appropriate choice 
for super-string computations would be determined by Wick rotation 
from the corresponding Euclidean theory. This suggestion relates the 
study of open string tachyon condensation to the so-called boundary 
Sine-Gordon model
$$ S_{BSG}[X]  \ = \ \frac{1}{4\pi} \int_\Sigma d^2 z\,  
    \partial X \bar \partial X + \int_{\partial \Sigma} du 
    \, \lambda \cos X(u) 
$$ 
in which $X = i X^{0}$ is a field with space-like rather than 
time-like signature. All spatial coordinates $X^\mu, \mu \neq 0,$ 
have been suppressed here since their contribution to the model 
is not affected by the interaction. 
\smallskip

The boundary Sine-Gordon theory has been studied rather intensively,  
see e.g. \cite{Callan:1994ub,Polchinski:1994my,Fendley:1994rh,
Recknagel:1998ih,Gaberdiel:2001zq}. Let us briefly review some 
of the most important results. To begin with, we point out 
that the boundary interaction is exactly marginal so 
that the theory is conformal for all values of $\lambda$. Properties 
of the model, however, depend crucially on the strength $\lambda$ 
of the boundary potential. In fact, it is well known that 
variations of $\lambda$ allow to interpolate smoothly between 
Neumann and Dirichlet boundary conditions. The former appear
for all integer values of $\lambda$ (in particular for $\lambda=0$, 
of course) while the latter are reached when $\lambda \in \frac12 + 
\QZ$. At these points, the theory describes a one-dimensional 
infinite lattice of D0 branes. From the geometric pictures we
can infer that the spectra of boundary excitations must also 
depend drastically on the parameter $\lambda$. In fact, at the 
points $\lambda \in \QZ$ with Neumann boundary conditions, the 
open string spectrum is continuous. If we now start tuning 
$\lambda$ away from these special values, the 
spectrum develops band gaps which become wider and 
wider until we reach the Dirichlet points $\lambda \in \frac12 
+ \QZ$ at which the spectrum is discrete. The first computation 
of the spectrum for generic values of $\lambda$ can be found in 
\cite{Polchinski:1994my} (see also \cite{Gaberdiel:2001zq} for 
a much more elegant derivation).    
\smallskip

Despite of these significant insights into the structure of the 
boundary Sine-Gordon model, there are several important quantities 
that remain unknown. This applies in particular to the boundary 
2- and 3-point functions and the bulk-boundary operator product 
expansions. In the string theoretic context, these missing data
determine the time-dependence of open string couplings on a 
decaying brane and the back-reaction to the bulk geometry. Our
desire to understand such important quantities is one of the 
main motivations to study the following, closely related 
world-sheet theory \cite{Gutperle:2003xf}, 
\begin{equation} \label{sgact}
 S_{TBL}[X] \ = \ \frac{1}{4\pi} \int_\Sigma 
                 d^2 z \, \eta_{\mu\nu} \partial X^\mu
                 \bar \partial X^\nu + \int_{\partial \Sigma} 
                 d u \, \mu_B \exp X^0(u) \ \ . 
\end{equation}  
This model has been named time-like boundary Liouville theory. Since the
tachyon vanishes in the far past, the model seems to describe a 
half-brane, i.e. a brane that dissolves as time progresses. After 
Wick rotation $X = iX_{0}$, the theory of the time coordinate
becomes
\footnote{Let us remark in passing that this model also arises 
in the context of flux line pinning in planar superconductors, 
see appendix F of \cite{Affleck:2004hz}} 
\footnote{In writing this action we think of the corresponding 
model as being defined through analytic continuation in the 
parameter $b$. This might be relevant for comparisons with 
previous results in models with exponential interactions.}    
\begin{equation}\label{BLact}  
 S_{BL}[X]  \ = \ \left( \frac{1}{4\pi} \int_\Sigma d^2 z\,  
    \partial X \bar \partial X + \mu \exp 2b X(z,\bz) + 
     \int_{\partial \Sigma} du \, \mu_B \exp b X(u)\right)^{\mu=0}_{b=i} 
    \ \ .  
\end{equation} 
We have written the interaction term for general parameters 
$b$ and also added a similar interaction in the bulk, mainly to 
emphasize the relation with boundary Liouville theory. What makes 
this relation so valuable for us is the fact that boundary Liouville 
theory has been solved over the last years \cite{Dorn:1994xn,
Zamolodchikov:1996aa,Ponsot:1999uf,Teschner:2001rv,Teschner:2003en,
Fateev:2000ik,Teschner:2000md,Hosomichi:2001xc,Ponsot:2001ng,
Zamolodchikov:2001ah}. Needless to stress that the solution includes 
explicit formulas for the bulk-boundary structure constants 
\cite{Hosomichi:2001xc} and the boundary 3-point couplings 
\cite{Ponsot:2001ng}. 

There is one crucial difference between Liouville theory and 
the model we are interested in: whereas the usual Liouville model 
is defined for real $b$, our application requires to set $b=i$. A 
continuation of results in Liouville theory from $c = 13 + 6b^2+
6b^{-2}\geq 25$ to $c=1$ (i.e.\ $b=i$) might appear to be a rather 
daring project, even more so as a naive inspection of the 
world-sheet action would suggest the $c=1$ model could not possibly 
be unitary. Nevertheless, we shall show below that the theory is 
entirely well-defined and unitary. For the pure bulk theory, a 
similar result was established in \cite{Schomerus:2003vv}. It 
was shown there that the bulk 3-point couplings of Liouville theory 
possess a $b=i$ limit which is well-defined for real momenta of the 
participating closed strings. The $b=i$ theory, however, is no longer 
analytic in the momenta. Furthermore, the limit turned out to agree 
with the $c=1$ limit of unitary minimal models which was constructed 
by Runkel and Watts in \cite{Runkel:2001ng}. Our analysis here 
extends these findings to the boundary theory. We shall see that 
the latter is considerably richer than the bulk model. 
\smallskip 

Some qualitative properties of the $c=1$ boundary model 
can be understood using no more than a few general observations. To 
begin with, let us reconsider the simpler bulk model. Recall from 
ordinary Liouville theory that it has a trivial dependence on the 
coupling constant $\mu$. Since any changes in the coupling can be 
absorbed in a shift of the zero mode, one cannot vary the strength 
of the interaction. This feature of Liouville theory persists when 
the parameter $b$ moves away from the real axis into the complex 
plane. As we reach the point $b=i$, our model seems to change 
quite drastically: at this point, the `Liouville wall' disappears 
and the potential becomes periodic. Standard intuition therefore
suggests that the spectrum of closed string modes develops gaps 
at $b=i$. But since the strength of the interaction cannot be 
tuned in the bulk theory, the band gaps must be point-like. 
Though our argument here was based on properties of the classical 
action which we cannot fully trust, the point-like band-gaps are 
indeed a characteristic property of the $c=1$ bulk theory 
\cite{Runkel:2001ng}. 
These band-gaps also explain why the couplings of Liouville theory 
cease to be analytic when we reach $c=1$. 
\smallskip 
 
For the boundary theory, we can go through a similar argument, but 
the consequences are more pronounced. In the presence of a 
boundary, Liouville theory contains a second coupling constant 
$\mu_B$ which controls the strength of an exponential interaction 
on the boundary of the world-sheet and is a real parameter of the 
model. In fact, the freedom of shifting the zero mode can only be
used for one of the couplings $\mu$ or $\mu_B$. Once more, the 
boundary potential becomes periodic at $b=i$ and hence the open 
string spectrum should develop gaps, as in the case of the bulk 
model. But this time, the width of these gaps can be tuned by 
changes of the parameter $\mu_B$. Hence, we expect the boundary 
Liouville theory to possess gaps of finite width at $c=1$. Our
exact analysis will confirm this outcome of the discussion. 
\smallskip 

After these more qualitative remarks on the model we are about 
to construct, we shall summarize our main results in more detail. 
In section 3 we shall provide the boundary states for a family of 
boundary conditions of the $c=1$ Liouville theory that is 
parametrized by one parameter $s \in \QR$. Formulas for these 
boundary states are trivially obtained from the corresponding 
expressions in $c \geq 25$ Liouville theory and, as we shall 
also explain, through particular limits of boundary theories in 
minimal models. We shall then use both approaches to establish the 
existence of finite band gaps in the boundary spectrum of the $c=1$ 
boundary theories. More precisely, we show that boundary 
fields $\Psi_p^{ss}$ in the boundary theory with parameter $s$ 
carry a momentum whose allowed values are taken from the following 
set 
$$ 
\S_s \ :=\ \{ p \in \QR\, | \, |\sin\pi p| \leq |\sin \pi s|\, \} \ \ .
$$ 
When $s$ is a half-integer, the set $\S_s$ fills the entire real 
line. As we vary $s$ away from such half-integer values, the 
allowed momenta are restricted to intervals of decreasing width 
until we reach integer values of $s$ where the spectrum of boundary 
fields becomes discrete. In this sense the parameter $s$ interpolates 
between Neumann-type and Dirichlet-type boundary conditions. Let us 
note that the boundary conditions with $s \in \QZ$ are related to 
those constructed by Runkel and Watts in \cite{Runkel:2001ng}. All 
other boundary theories, however, are new. 

It is worthwhile pointing out how closely the $c=1$ limit of Liouville 
theory mimics the behavior of the boundary Sine-Gordon model. In both 
theories we can interpolate smoothly between continuous and discrete 
boundary spectra. Even the geometric interpretation of the Dirichlet 
points is very similar: in the context of the $c=1$ Liouville model 
they describe a semi-infinite array of point-like branes from which 
individual branes can be removed through shifts of the boundary 
parameter $s$. Let us point out that finite arrays of such point-like
branes are described by the $c=1$ limit of ZZ branes.\footnote{The 
two integers $(m,n)$ that parametrize the ZZ branes correspond to 
the transverse position and the length of the finite array.} Hence, 
the picture we suggest for the Dirichlet points of the $c=1$ FZZT 
branes provides a nice geometric interpretation for the well-known 
relation of FZZT and ZZ branes in Liouville theory 
\cite{Martinec:2003ka,Teschner:2003qk}.   
\smallskip 

There are two important quantities that we shall construct for
all these new boundary theories. These are the boundary 2-point 
function and the bulk-boundary 2-point function. For the former 
we obtain
\begin{eqnarray}
\langle \Psi^{s_1 s_2}_{p_1} (u_1) \Psi^{s_2s_1}_{p_2}(u_2) 
  \rangle^{c=1} & = & \frac{{\cal{N}}^{s_1s_2}(p)}{|u_1-u_2|^{2p^2_1}} 
  \left[ 
 \delta(p_1+p_2) + \R^{s_1s_2}_{c=1}(p_1) \delta(p_1-p_2)
 \right] \\[3mm] 
\mbox{ where } & & \R^{s_1s_2}_{c=1}(p) \ = \ 
    \R^{is_1\ is_2}_{c=25} (1-p)^{-1}\ \  .     
\end{eqnarray}      
Detailed formulas for ${\cal N}$ (see eq.\ (\ref{2ptfctres1})) 
and for the so-called reflection amplitude $\R$ are spelled out in
section 4. Here, we have expressed the result in terms of the
reflection amplitude for the theory at $b=1$. It is quite remarkable
that the reflection amplitudes of the two models are related in such a
simple way. Let us point out, though, that the boundary reflection
amplitude for the $c=1$ theory is only defined within a subset
$\S_{s_1s_2}$ of momenta $p$.   
\smallskip 

It is more difficult to write down the bulk-boundary coupling
$B$ of the $c=1$ model, i.e.\ the coefficient in the 2-point 
function of a bulk field $\Phi$ with one of the allowed 
boundary fields, 
$$ \langle \Phi_{p_\a}(z,\bz) \Psi^{ss}_{p_\b}(u)\rangle^{c=1}
 \ = \ \frac{B(p_\a,p_\b,s)}{|z-\bz|^{2p_\a^2 -p_\b^2}\,  
 |z-u|^{2p_\b^2}}  \ \ . 
$$
We shall prove below that this coupling $B$ is given by  
\begin{eqnarray} \label{Bcoupl} 
 B_{c=1} (p_\a,p_\b,s) & = & 
4\pi^{2} i (\pi \mu_{\text ren})^{-p_\a-p_\b/2} \, 
    e^{- i\pi s (p_{\beta}+1) }\,\Gamma (-2p_{\beta })^{-1}\, 
    \times \\[2mm]
& &  \hspace*{-3.5cm}\nonumber
\times \,\bigg(  \frac{e^{-2\pi ip_{\alpha} (s+1)}}{1-e^{-4\pi
ip_{\alpha }}} \frac{g(y_0)^2 e^{-\frac{ih (y_{0})}{2\pi}}}{y_0^{2} \, 
h''(y_0)}\, \exp \int_0^\infty \frac{dt}{t} \ \frac{\left(e^{-p_\b t} 
   - e^{2p_\a t}\right)
\sinh^2\frac{p_{\b}t}{2}}{\sinh^2\frac{t}{2}}-\ (p_{\alpha}\rightarrow
-p_{\alpha }) \bigg) 
\end{eqnarray} 
where $p_\b \in \S_s$ and the functions $g$ and $h$ are defined through 
\begin{eqnarray} 
g(y)& =& \left(\frac{1-A_0}{1-A_0y}
\right)^{-p_{\alpha}-p_{\beta}/2}
\left(\frac{1-B_0}{1-B_0y} \right)^{-p_{\beta}/2}
\left(\frac{1-C_0}{1-C_0y} \right)^{p_{\alpha}+1/2}
\frac{y^{\frac{1+s-p_{\beta}}{2}}}{\sqrt{1-y}} \\[4mm]
h (y)&=& \Li (A_{0})-\Li (A_{0}y)+\Li (B_{0})-\Li (B_{0}y)\nonumber \\
&&-\Li (C_{0})+\Li (C_{0}y)-\Li (1)+\Li (y)+ \log y \log z_{0} \ \ .
\end{eqnarray} 
Here, $\Li $ denotes the dilogarithm and 
we have abbreviated $B_0 = \exp (-2\pi i p_\b), C_0 = 
\exp(-4\pi i p_\a), A_0 = B_0 C_0, z_{0}=\exp (-2\pi i (s-p_{\beta }))$  
and the  parameter $y_0= 
y_0(p_\a,p_\b,s)$ is one of the two solutions of the following 
quadratic equation    
$$  \exp(y\, h' (y)) \ = \ 1 \ \ . $$ 
For more details and explicit formulas, see section 5. We also show 
that the coefficients in the bulk-boundary operator product expansion
vanish whenever the open string momentum $p_\beta$ lies within the 
band gaps, i.e.\ when $p_\b \not \in \S_{s}$. This provides another 
non-trivial consistency check for the couplings we propose. Let us 
add that correlators with boundary insertions in a similar model 
where also discussed recently in \cite{Kristjansson:2004mf}. The 
techniques used there, however, only allowed to determine such 
correlations functions for a discrete set of boundary momenta 
$p_\b$.

\section{The $c=1$ limit of the bulk theory} 
\setcounter{equation}{0} 

In this section we shall mainly review some results on the $c=1$ limit 
of bulk Liouville theory \cite{Schomerus:2003vv} (see also 
\cite{Strominger:2003fn}).  Our derivation of the 3-point couplings, 
however, is different from the one given in \cite{Schomerus:2003vv}. 
The new construction is simpler and uses some of the same techniques 
that we shall also employ to analyze the boundary model later on. 
\smallskip 
   
To begin with, let us recall a few standard facts about the usual 
$c\geq 25$ Liouville theory. As in any bulk conformal field theory, 
the exact solution of the Liouville model is entirely determined by 
the structure constants of the 3-point functions for the (normalizable) 
primary fields 
\begin{equation}\label{flds} 
 \Phi_\a(z,\bz) \ \sim \ e^{2 \a X(z,\bz)} 
\ \ \ \mbox{ with } \ \ \a \ = \ \frac{Q_b}{2} + i p
 \ \ \ , \ \ \ p\ \geq \ 0 \ \  \end{equation}
where $Q_b = b + b^{-1}, b \in \QR$. These fields are primaries 
with conformal weight $h_\a = \bar h_\a= \a (Q_b-\a)$ under the action 
of the two Virasoro algebras whose central charge is $c = 1 + 6 Q^2_b$. 
For real values of $b$, the couplings of three such fields are given 
by the following expression 
\begin{equation}\label{bulktpf}
C_b (\alpha_{1},\alpha_{2},\alpha_{3})\ =\ 
\big(\pi \mu \gamma (b^{2})b^{2-2b^{2}} \big)^{(Q_{b}-2\tilde{\alpha})/b}
\frac{\Upsilon' (0|b)}{\Upsilon (2\tilde{\alpha}-Q_b|b)}
\prod _{j=1}^{3}\frac{\Upsilon (2\alpha_{j}|b)}{\Upsilon (2\tilde{\alpha}_{j}|b)}
\end{equation}
where $\ta$ and $\ta_j$ are linear combinations of $\a_j$, 
\begin{align*}
\tilde{\alpha}\ =&\ \frac{1}{2} (\alpha_{1}+\alpha_{2}+\alpha_{3}) &
\tilde{\alpha}_{j}\ =&\ \tilde{\alpha}-\alpha_{j} \ \ , 
\end{align*}
$\gamma(y) = \Gamma(y)/\Gamma(1-y)$ is a quotient of ordinary 
$\Gamma$-functions and the function $\Y$ is defined in terms of 
Barnes' double Gamma function $\Gamma_2$ (see appendix A.2) by 
\begin{equation} \label{ZY} 
 \Y(\a|b) \ := \ \Gamma_2(\a|b)^{-1}\, 
                    \Gamma_2(Q_b-\a|b)^{-1} \ \ . 
\end{equation}  
The solution (\ref{bulktpf}) was first proposed several years ago by H.\ Dorn 
and H.J.\ Otto \cite{Dorn:1994xn} and by A.\ and Al.\ Zamolodchikov 
\cite{Zamolodchikov:1996aa}. Crossing symmetry of the conjectured 
3-point function was then checked analytically in two steps by 
Ponsot and Teschner \cite{Ponsot:1999uf} and by Teschner 
\cite{Teschner:2001rv,Teschner:2003en}.  
\smallskip 

It is well known that Barnes' double Gamma function is analytic 
for $\Re b \neq 0$. In the limit $b=i$, however, the function 
becomes singular. Nevertheless, the combinations of double Gamma
functions that appear in the various couplings of Liouville theory 
turn out to be well defined, though they are no longer analytic. 
Most of our analysis of the $b=i$ limit is based on the following 
relation between double Gamma functions with parameter $b$ and 
$-ib$,  
\begin{equation} \label{dGdGrel} 
 \Gamma_2(ip + n \frac{Q_{b}}{2}|b) \ = \ 
  \frac{(-q;q^2)_\infty e^{\frac{i\pi}{4} (ip + (n-1)
  \frac{Q_{b}}{2})^2}}{(e^{2\pi b^{-1} p} (-1)^n q^n;q^2)_\infty} 
   \ \Gamma_2^{-1}(-ib - p + i n \frac{Q_{b}}{2}|-ib) 
\end{equation} 
which holds for $\Im b^2 > 0$. Here we have introduced $q = 
\exp(-i \pi/b^2)$ and the 
so-called q-Pochhammer symbols $(.;.)_\infty$ (see appendix A.1). 
Relation (\ref{dGdGrel}) follows from a rotation of the contour in 
the integral representation (\ref{BarnesG}) of the double Gamma function. 
The factor involving the q-Pochhammer symbols assembles contributions 
from the poles in the integrand of the double Gamma function (see 
appendix A.2 for details). Using formula (\ref{pochasymp}) for the 
behavior of our q-Pochhammer symbols at $b=i$ we conclude 
\begin{eqnarray} 
 \Gamma_2(ip + n \frac{Q_{b}}{2}|b) & \sim & \sqrt{2} \, 
    e^{- i \frac{\pi}{4}p^2} \, \left( 2 |\sin \pi p|
     e^{-\pi i (p - [p]-1/2)} \right)^{-(p+1-n)/2} \, \times \\[2mm] 
  & & \hspace*{2cm} \times \,  e^{\frac{1}{\e} 
   \left( Li_2(e^{-2\pi i p}) - Li_2(1)\right)}  \Gamma_2^{-1}(1-p|1) \, 
     \left(1 + o(\e^{0})\right) \label{Gasym} \nonumber
\end{eqnarray} 
with $\epsilon = 2 \pi i b^{-1} Q_{b}$ and $[p]$ denoting the 
integer part of $p$.
Note that the double Gamma function diverges as we send $b 
\rightarrow i$ (or $\epsilon \rightarrow 0$) with a divergent 
factor involving the Dilogarithm $Li_2$. 
\smallskip 

In the expression for closed string couplings, Barnes' double Gamma
functions appears only through $\Y$, and using a standard dilogarithm 
identity (see appendix A.3) one may show that                
\begin{equation}\label{upsilonasymp}
\Upsilon (ip+ n \frac{Q_{b}}{2}|b)\ \stackrel{b \to i}{\sim}\
\frac{1}{2} \, e^{-\frac{1}{\epsilon}\big(\lambda (p)-3\Li (1)\big)}
\, e^{i\pi p^{2}/2}\, e^{-i\pi (p - n + 1)(p-[p]-1/2) }\,  
    \Upsilon (1-p|1)^{-1} \ \ .
\end{equation}
Here, $\lambda (p)$ is a continuous periodic function which is 
quadratic on each interval $[n,n+1]$ for $n\in \mathbb{Z}$,
\[
\lambda (p)\ =\ 2\pi^{2} (p-[p]-1/2)^{2}
\ \ .
\]
Observe that the divergent factor in $\Y$ is much simpler than for 
the double Gamma functions it is built from. In the function $\Y$, 
the divergence comes from a product of q-Pochhammer symbols. The 
latter is closely related to Jacobi's function $\vartheta_1$ and 
its divergence may be controlled more directly using modular 
properties of $\vartheta$-functions (see \cite{Schomerus:2003vv}). 
\smallskip 

Using once more the formula (\ref{dGdGrel}) and some simple facts 
about Barnes' double Gamma function is is not difficult to show that 
\begin{equation}\label{upsilonprimeasymp}
\Upsilon '(0|b)\ \stackrel{b\to i}{\sim} \ \frac{2\pi^{2}}{\epsilon}
\Upsilon (1|1)^{-1} (1+o (\epsilon^{0}))\ \ .
\end{equation}
With the help of the two asymptotic expressions~\eqref{upsilonasymp} 
and~\eqref{upsilonprimeasymp}, we find
\begin{align}
\nonumber
\frac{\Upsilon' (0|b)}{\Upsilon (2\tilde{\alpha}-Q_{b}|b)} &\prod
_{j=1}^{3}\frac{\Upsilon (2\alpha_{j}|b)}{\Upsilon
(2\tilde{\alpha}_{j}|b)} \ = \ \frac{\Upsilon' (0|b)}{\Upsilon
(Q_{b}/2+2i\tilde{p}|b)} \prod
_{j=1}^{3}\frac{\Upsilon (Q_{b}+2ip_{j}|b)}{\Upsilon
(Q_{b}/2+2i\tilde{p}_{j}|b)}\\
\nonumber
= &\  e^{-\frac{1}{\epsilon} F
(p_{1},p_{2},p_{3})}\, \frac{4\pi^{2}}{\epsilon}\   
e^{2 \pi i \tilde{p} (\tilde{p}-[2\tilde{p}]-1/2)}
\prod _{j=1}^{3} \frac{e^{\pi i p_{j}}e^{-\pi i (2p_{j}-1)
(p_{j}-[2p_{j}]-1/2)}}{e^{-\pi i 2\tilde{p}_{j}
(\tilde{p}_{j}-[2\tilde{p}_{j}]-1/2)}}\\
&\ \times \ \frac{\Upsilon (1-2\tilde{p}|1)}{\Upsilon (1|1)}
\prod _{j=1}^{3}\frac{\Upsilon(1-2\tilde{p}_{j}|1)}
{\Upsilon (1-2p_{j}|1)}\  (1+o (\epsilon^{0}))
\label{bulktpfasymp}
\end{align}
where
\begin{equation}\label{defofF}
F (p _{1},p_{2},p_{3})\ =\ \sum_{j=1}^{3}\lambda (2p_{j})
-\sum_{j=1}^{3}\lambda (2\tilde{p}_{j}) -\lambda (2\tilde{p}) +3\Li (1)
\ \ .
\end{equation}
It can be shown that $F (p_{1},p_{2},p_{3})\geq 0$, so that the
exponential is never diverging, but it can suppress the whole
three-point coupling. We write $2p_{j}=[2p_{j}]+f_{j}$ and find that
$F$ is zero if either 
\[
[2p_{1}]+[2p_{2}]+[2p_{3}]\ \ \text{even, and } \ |f_{1}-f_{2}|\leq
f_{3}\leq \min \{f_{1}+f_{2},2-f_{1}-f_{2} \}\ \ , 
\]
or
\[
[2p_{2}]+[2p_{2}]+[2p_{3}]\ \ \text{odd, and } \ |f_{1}-f_{2}|\leq 1-f_{3}\leq
\min \{ f_{1}+f_{2},2-f_{1}-f_{2}\} \ \ ,
\]
otherwise $F$ is strictly greater than zero and the three-point
coupling vanishes for $\epsilon\to 0$. In more formal terms, the 
function 
$$ P(2p_{1},2p_{2},2p_{3}) \ := \ \lim_{\epsilon \rightarrow 0} 
     e^{-\frac{1}{\epsilon} F(p_1,p_2,p_3)} $$ 
assumes the value $P=1$ when one of the above conditions is fulfilled 
and it vanishes otherwise.  With the new function $P$ being introduced, 
we can use eq.~\eqref{bulktpfasymp} to recast the 3-point couplings in 
the form 
\begin{eqnarray}
C_{c=1}(p_1,p_2,p_3) & \stackrel{b\to i}{\sim} & 
 (\pi \mu_{ren})^{-2\tilde p} \, \frac{4\pi^{2}i}{\epsilon}\  
  \prod _{j=1}^{3}\, e^{-\pi i[2p_{j}]} \, 
 P (2p_{1},2p_{2},2p_{3}) \times \nonumber\\[2mm] 
   & & \hspace*{3cm} \times \ 
 \frac{\Upsilon (1-2\tilde{p}|1)}{\Upsilon (1|1)}
     \prod _{j=1}^{3}\frac{\Upsilon
 (1-2\tilde{p}_{j}|1)}{\Upsilon (1-2p_{j}|1)} \ \ . 
\label{bulk3ptfunc}
\end{eqnarray}
In our conventions, $Q_{b}\sim \epsilon/2\pi$, so up to a factor $i$ we
reproduce precisely the result of~\cite{Schomerus:2003vv} (see eqs.\
(4.8) and (5.3) of that article). To obtain a finite limit, the bulk
fields should be renormalized by a factor $\epsilon$ and correlators
on the sphere receive an additional factor $\frac{1}{\epsilon^{2}}$
due to a rescaling of the vacuum (see section 4.2 in
\cite{Runkel:2001ng} for a related discussion). This prescription
produces a finite 3-point function and it does not affect the 2-point
function.
\smallskip 

At first, the form of the couplings, and in particular the non-analytic
factor $P$, may seem a bit surprising. It is therefore reassuring that 
exactly the same couplings emerged several years ago through a $c=1$ 
limit of unitary minimal models \cite{Runkel:2001ng}. Runkel and Watts 
investigated the consistency of these couplings and demonstrated that 
crossing symmetry holds when all the half-integer momenta $p$ are 
removed from the spectrum. As we have argued in the introduction, the 
appearance of such point-like band-gaps is rather natural from the 
point of view of Liouville theory. 

\section{The $c=1$ limit of the boundary theory} 
\setcounter{equation}{0} 

Now we are ready to analyze the boundary model. We shall begin with 
the so-called FZZT boundary conditions of Liouville theory. They 
describe extended branes with an exponential potential for open strings 
(see eq.\ (\ref{BLact})). Taking the $c=1$ limit of their 1-point 
function is straightforward and results in an expression with analytic 
dependence on the closed string momentum. These $c=1$ boundary states 
are parametrized through one real parameter $s$ and for non-integer 
values of $s$ they can be reproduced by taking an appropriate limit 
of boundary conditions in unitary minimal models. The latter 
construction will provide a first derivation of the spectrum of 
boundary vertex operators, including the precise position and width 
of the band-gaps. 

When $s \in \QZ$, the bands become point-like. At these special values
of the parameter $s$, our two constructions through Liouville theory 
and minimal models do not produce the same boundary states, though 
they still give closely related results. From minimal models we 
obtain the discrete set of boundary conditions that is already 
contained in the work of Runkel and Watts. These agree with the 
$c=1$ limit of ZZ branes in Liouville theory and they possess a 
simple geometric interpretation as finite arrays of point-like 
branes. The relation between ZZ and FZZT branes in Liouville 
theory then implies that we can interpret FZZT branes with 
$s \in \QZ$ as half-infinite arrays of point-like branes in 
the $c=1$ model. 
\medskip

\subsection{Boundary states of FZZT branes at $c=1$} 

It is well known that for real $b$, Liouville theory admits a 
1-parameter family of boundary conditions which correspond to 
branes stretching out along the real line. The boundary 
state of this theory was found in \cite{Fateev:2000ik}, 
\begin{eqnarray} 
\langle \Phi_\a(z,\bz)\rangle_s & = & \Gamma(1+2ibp)\,  
     \Gamma(1+2ip/b) \,  \cos(2\pi s p) \, 
      \frac{(\pi \mu \gamma(b^2))^{-ip/b}}{2\pi i 2^{1/4} p}
      \, \frac{1}{|z-\bz|^{2h_\a}}\\[3mm] 
   & &  \mbox{where} \ \ \  \mu_B^2 \sin \pi b^2 \ = \ 
           \mu \cosh^2 \pi s b   \nonumber  
\end{eqnarray}
and $\a = Q_{b}/2 + ip$, as usual. Note that the coupling on the 
right hand side is analytic in $b$ and hence there is no 
problem to extend the FZZT brane boundary state into the 
$c=1$ theory,
\begin{equation} \label{1ptfceq1} 
\langle \Phi_\a(z,\bz)\rangle^{c=1}_s \ = \   
      \frac{(\pi \mu_{\rm ren})^{-p}}{ i 2^{1/4}} 
     \frac{\cos 2\pi s p }{\sin 2\pi p} \, 
       \, \frac{1}{|z-\bz|^{2p^2}}\ \ . 
\end{equation}
This boundary state is related to its Lorentzian analogue 
\cite{Larsen:2002wc,Gutperle:2003xf} by analytic continuation. 
It is, however, not entirely obvious that eq.\ (\ref{1ptfceq1}) 
really provides the one-point function of a consistent boundary 
conformal field theory. We shall argue below that this is the case, 
at least as long as the parameter $s$ remains real. 
\smallskip 

Let us now see how such a boundary state of the $c=1$ 
model can arise from a limit of boundary minimal models. To 
begin with, we need to set up a few notations. We shall label 
minimal models by some integer $m=3,4,5,\dots$ so that $c 
= 1-6/m(m+1)$. For representations of the corresponding Virasoro 
algebra we use the Kac labels $(r,r')$ with $1\leq r \leq m-1$ and 
$1 \leq r' \leq m$ along with the usual identification  $(r,r') \sim 
(m-r,m+1-r')$. Recall that the boundary conditions of the theory 
are in one to one correspondence with the representations, i.e.\ 
that they are also labeled by Kac labels $(\rho,\s)$. The 
associated one-point functions are given by 
\begin{equation} \label{1ptfMM} 
 \langle \Phi_{r,r'}(z,\bz)\rangle_{(\rho,\s)} \ = \ 
   \frac{2^{3/4}}{(m(m+1))^{1/4}}\,  \frac{\sin \pi \rho (r/t - r') 
           \sin \pi \s (r-r't)}{\sqrt{\sin (r/t - r') 
           \sin \pi (r-r't)}} \, \frac{1}{|z-\bz|^{2h_{r,r'}}}
\end{equation}  
where $t= m/(m+1)$ and $h_{r,r'}$ is the conformal dimension of
the primary field with Kac labels $(r,r')$, 
$$ h_{r,r'} \ = \ \frac{\left((mr'-(m+1)r)^2-1\right)}{4 m (m+1)} 
   \ =: \ p_{r,r'}^2 \ \ . $$ 
Here, we have introduced the new quantity $p_{r,r'}$ that can be 
regarded as a discrete  analogue of the continuous momentum $p$  
when $c < 1$. 
\smallskip 

In taking the limit $m \rightarrow \infty$, the bulk fields 
$\Phi_{r,r'}$ of minimal models approach the fields $\Phi_p$ 
in the continuum theory. The momentum variable $p$ is related to the
Kac-labels by  
$$ 2 p \ = \ 2 p_{r,r'} \ \sim \ r-r' + \frac{1}{2m+2}\, (r+r')
  + \mathcal{O}(1/m^2) \ \ .$$
In other words, the integer part of the quantity $2p$ is determined 
by the distance of the Kac label $(r,r')$ from the diagonal $r=r'$, 
whereas the fractional part $2p-[2p]$ corresponds to the (rescaled) 
position along the diagonal. For the A-series of minimal models, the 
bulk spectrum is obtained by filling every point of the Kac table 
above the diagonal. When we send $m$ to infinity, approximately the  
same number of fields appears at any given distance from the diagonal 
of the Kac table and these are distributed almost homogeneously along 
the diagonal $r' = r$. Hence, the spectrum  of $p$ in the 
$c=1$ limit is homogeneous. Only integer values of $2p$ are not part 
of this spectrum since the Kac labels satisfy $r,r' \geq 1$. This 
leads to the point-like band-gaps in the spectrum of bulk fields
that we also saw emerging from Liouville theory.     
\smallskip  

To obtain a one-parameter family of branes with 1-point function 
given by eq.\ (\ref{1ptfceq1}) we propose the following Ansatz, 
\begin{equation} \label{prescript}
 (\rho,\s) \ = \ (\, [f_s m] + [s] + 1 \, , \, [f_s m]+ 1 \, ) \ \ .
\end{equation} 
Here, $s \in \QR$ is the parameter of the resulting limiting 
boundary condition. $[s]$ and $f_s$ denote the integer and 
fractional part of $s$, respectively. The prescription 
(\ref{prescript}) instructs us to pick a boundary condition 
that is represented by a point at distance $[s]$ from the diagonal 
of the Kac table and to scale its projection to the diagonal with 
$m$. Our first aim now is to show that this recipe reproduces
the 1-point function (\ref{1ptfceq1}) in the $c=1$ Liouville 
theory. The main idea is to rewrite the product of $\sin$-functions 
in the numerator of eq.\ (\ref{1ptfMM}) as follows
\begin{eqnarray}  & & 2 \sin \pi ( [f_s m]+[s]+1) (r/t - r') \, 
   \sin \pi ([f_s m] +1)(r-r't) \nonumber \\[2mm] 
  & & \hspace*{3cm} \stackrel{m\rightarrow \infty}{\sim}  \ 
    \cos 2 \pi p s - 
  \cos 4 \pi p  ([f_s m]+1+[s]/2) + \dots \nonumber 
\end{eqnarray} 
with $p \sim (r-r't)/2$ being the momentum in the $c=1$ theory. 
Provided $s \not \in \QZ$, the second term in the previous 
formula oscillates rapidly when we send $m$ to infinity and 
hence we can drop it in the limit 
(recall that all correlation functions should be understood as 
distributions in momentum space). Together with the other factors 
in the 1-point function (\ref{1ptfMM}) of minimal models, a short 
analysis similar to the one carried out in \cite{Runkel:2001ng} 
gives 
$$ \langle \Phi_p(z,\bz)\rangle_s \ = \ 2^{-1/4} 
    \frac{\cos 2 \pi s p }{|\sin 2 \pi p|}\, 
     \frac{1}{|z-\bz|^{2p^2}} 
      \ \ . 
$$ 
The result agrees with the 1-point function (\ref{1ptfceq1}) of the 
$c=1$ Liouville model up to some prefactors which may be absorbed by 
an appropriate renormalization of the bulk fields. If, on the other 
hand, the label $s$ is an integer, then our limit of boundary 
conditions in minimal models becomes 
\begin{equation} \label{MML} 
 \langle \Phi_p(z,\bz)\rangle^{\rm MM}_s \ = \  
    \frac{\cos 2 \pi s p - \cos 2\pi (s+2) p}{2^{1/4} |\sin 2 \pi p| \ 
       |z-\bz|^{2p^2}} \ \sim \  \langle \Phi_p(z,\bz)\rangle_s
   -  \langle \Phi_p(z,\bz)\rangle_{s+2}\ \ . 
\end{equation} 
The last identification with the $c=1$ limit of boundary states in 
Liouville theory holds up to some factors that are due to the slightly 
different normalizations of bulk fields in minimal models and 
Liouville theory (see \cite{Schomerus:2003vv}). Our result (\ref{MML}) 
is very reminiscent of a similar relation between localized and extended 
branes in Liouville theory \cite{Martinec:2003ka,Teschner:2003qk} and we 
shall comment on the precise connection and its geometric interpretation 
in the $c=1$ model below.  
\smallskip 

Now that we have identified the construction of the boundary 
state from minimal models, it is instructive to study the 
associated spectrum of boundary fields. According to the usual 
rules \cite{Cardy:ir}, the boundary spectrum of the brane $(\rho,\s)$ 
contains fields with Kac labels $r = 1,3, \dots, \min(2\rho-1,
2m-2\rho-1)$ and $r' = 1,3, \dots, \min(2\s-1,2m-2\s+1)$. Using 
the identification $(r,r') \sim (m-r,m+1-r')$ we can assume that 
$r\geq r'$. Hence, the Kac labels of boundary fields fill every 
second lattice point within two triangles in the Kac table (see 
Figure 1). In the first triangle with corner at $(1,1)$, the 
difference $r-r'$ of Kac labels is even. For Kac labels in the 
second triangle, on the other hand, $r-r'$ is an odd integer 
since the reflection $(r,r') \rightarrow (m-r,m+1-r')$ shifts 
the difference of Kac labels by one unit. Recall that the 
quantity $r-r'$ measures the distance of a point in the Kac 
table from the diagonal and thereby determines the integer 
part of $2p$ in the $c=1$ theory. We conclude that the two 
triangles contribute boundary fields with $[2p]$ being even 
or odd integer, respectively. Furthermore, points from the 
first triangle are uniformly distributed in the direction 
along the diagonal up to a maximal value $0 <  (\min(2\rho-1,
2m-2\rho-1) + \min(2\s-1,2m-2\s+1))/2(m+1) < 1$. A similar 
observation holds true for the second triangle. After we 
have sent $m$ to infinity, this implies that the spectrum 
of $p$ is given by 
$$ \S_{s} \ = \ \{ \, p \in \QR \, | \, - \min(f_s,1-f_s)
    < p < \min(f_s,1-f_s) \mod \, 1 \ \} \ \ . $$ 
Hence, the momentum $p$ has bands of width $2\min(f_s,1-f_s)$ 
centered around integer momentum $p$. The gaps between these 
bands widen while we vary $s$ between a half-integer and an 
integer value. Once we reach a point $s\in \QZ$, the boundary 
spectrum becomes discrete.
\begin{figure} 
\centering
\scalebox{.5}{\epsfbox{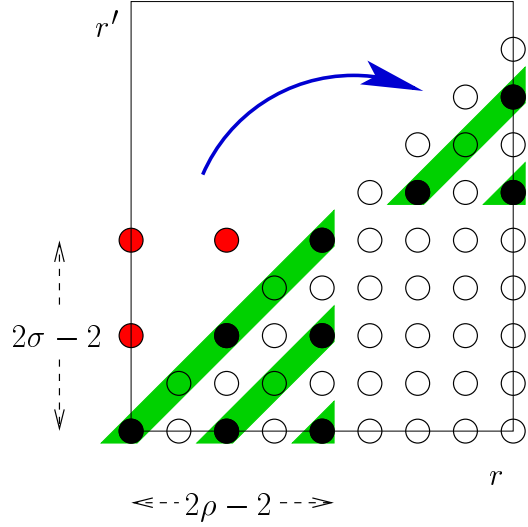}} 
\caption{Boundary excitations for the boundary condition 
$(\rho,\sigma)$ fill a rectangle, or, after reflection, 
two triangles in the Kac table. In the large $m$ limit, 
they form bands with band-gaps around $2p=1$.}  
\end{figure}
\smallskip 

\subsection{ZZ and FZZT branes at the Dirichlet points} 

Branes in the $c=1$ limit of minimal models with a discrete 
open string spectrum have also been constructed by Runkel and 
Watts (see also~\cite{Graham:2001tg}). Actually, the branes that we
obtained through the limit of minimal models when $s\in \QZ$ 
are identical to the limit of $(s+1,1)$ branes in minimal models
and hence to the discrete boundary conditions that were found
by Runkel and Watts. We wish to point out that these branes 
may also be obtained from the so-called ZZ branes of Liouville 
theory. In general, the ZZ branes are labeled by a pair $(m,n)$ 
of positive integers and they possess the following 1-point 
functions, 
\begin{equation*}
\langle \Phi_\a(z,\bz)\rangle_{(m,n)}\ = \ 
-\frac{2^{3/4}2\pi ip}{\Gamma (1-2ibp)\Gamma (1-2ip/b)}\, \frac{\sinh 2\pi
mp/b\, \sinh 2\pi nbp}{\sinh 2\pi p/b\, \sinh 2\pi bp}\,  
\frac{(\pi \mu \gamma (b^{2}))^{-ip/b}}{|z-\bz|^{2h_\a}} \ \ .
\end{equation*}
If we send the parameter $b$ to $b=i$, these 1-point functions
assume the following form 
\begin{equation*}
\langle \Phi_p(z,\bz)\rangle^{c=1}_{(m,n)} \ = \ 
- i(\pi \mu_{\ren})^{-p}2^{3/4}\frac{\sin 2\pi mp\, \sin 2\pi np}{\sin
2\pi p}\, \frac{1}{|z-\bar{z}|^{2p^{2}}}\ \ .
\end{equation*}
We can rewrite the coupling on the right hand side with the 
help of the trigonometric identity 
$$  \frac{\sin 2\pi m p \ \sin 2\pi n p}{\sin 2 \pi p }
   \ = \ \sum_{l=0}^{\min(m,n)-1} \, \sin 2\pi p (m+n -2l-1)   
$$ 
to read off the following geometric interpretation of the 
ZZ brane with label $(m,n)$ in the $c=1$ theory: it corresponds
to an array of $\min(m,n)$ point-like branes that are distributed 
equidistantly such that the rightmost brane is located at $m+n$. 
This interpretation agrees with the geometry of branes that 
emerges from the coset construction of minimal models (see 
\cite{FreSch}). 
\smallskip 

With this in mind, we shall also be able to interpret our FZZT 
branes at the Dirichlet points $s \in \QZ$. Let us recall that 
the $c=1$ limit of FZZT branes with integer label $s$ differs 
from the limit of minimal model branes. The precise relation 
between their two boundary states is given in eq.\ (\ref{MML}). 
On its right hand side we consider the difference of one-point 
couplings for two FZZT branes whose labels differ by two units. 
This is claimed to be the same as the one-point coupling to the 
discrete brane with label $(s+1,1)$. Therefore, the brane 
configuration that is associated to the FZZT brane with 
integer label $s$ contains an additional point-like object 
with transverse position parameter $s$ along the real line. 
The latter gets removed as we shift from $s$ to $s+2$. Our 
observation suggests to think of the $c=1$ FZZT branes with 
integer $s$ as being built up from discrete objects, or more 
specifically, as some half-infinite array of point-like 
objects. They occupy points with even or odd integer position 
$x$ in target space, depending on whether $s$ is odd or even. 
Changes of $s$ by one unit should therefore correspond to a 
simple translation of the entire array in the target space.   

At least qualitatively, such an interpretation in terms of 
point-like branes appears 
consistent with our previous study of brane spectra on FZZT 
branes. In fact, we noticed before that their bands shrink 
to points when $s$ reaches integer values. Let us also stress 
that our geometric picture of the branes with integer $s$ is 
very similar to the interpretation of the Dirichlet points 
in the boundary Sine-Gordon model (see introduction). 
\smallskip 

More support for the proposed geometric pictures comes from 
considering spectra involving the ZZ branes in the $c=1$ 
theory. The annulus amplitude for two ZZ branes with labels 
$(m,n)$ and $(m',n')$ was found in \cite{Zamolodchikov:2001ah} 
to be of the form 
\begin{eqnarray*}
Z_{(m,n),(m',n')}(q) & = & \sum_{k=0}^{\min(m,m')-1} 
   \ \sum_{l=0}^{\min(n,n')-1} \ \chi_{m+m'-2k-1,n+n'-2l-1}(q)   
  \\[2mm] 
 \mbox{where} \ \ \chi_{m,n}(q) & = & \eta^{-1}(q) \, 
   \left( q^{-(m/b+nb)^2/4}-q^{-(m/b-nb)^2/4}\right) \\[2mm]
  & \stackrel{b=i}{=} &   \eta^{-1}(q) \, \left( 
     q^{(m-n)^2/4}-q^{(m+n)^2/4}\right)\ \ . 
\end{eqnarray*}  
Note that in the $c=1$ theory, excitations of the ZZ branes
have non-negative conformal dimensions. Since $\chi_{m,n}$ 
are characters of irreducible Virasoro representations at 
$c=1$, the expression for the annulus amplitude coincides 
with what is obtained from the limit of Cardy type 
branes in unitary minimal models. A particularly simple 
case arises from setting $(m,n) = (1,1)$ and $(m',n') =
(r,r)$, 
\begin{equation}  \label{ann1l} 
Z_{(1,1),(r,r)}(q) \ = \ \chi_{r,r}(q) \ =\  
    \eta^{-1}(q)\,  \left( 1-q^{r^2}\right)\ \ .  
\end{equation} 
According to our previous discussion, this amplitude encodes
the spectrum of open strings between a single point-like 
brane and an array of length $2r$.    
\smallskip 

For the annulus amplitude between a ZZ brane $(m,n)$ and an 
FZZT brane at parameter $s$ one has
\begin{eqnarray} \label{ZZFZZT}
Z_{(m,n),s}(q) & = & \sideset{}{'}\sum_{k=1-m}^{m-1} \  
                     \sideset{}{'}\sum_{l=1-n}^{n-1} \ 
               \chi_{(s+i(k/b+lb))/2}(q) \\[2mm]
\mbox{where} \ \ \chi_p(q) & = & \eta^{-1}(q) \, q^{p^2}\ \ . 
\nonumber 
\end{eqnarray}
Here, $\Sigma'$ denotes a summation in steps of two. Let us 
point out that once more, the open string excitations between 
ZZ and FZZT branes have real, non-negative conformal dimensions 
when $b=i$. This is in contrast to the situation for $c \geq 25$ 
where the spectrum contains fields with complex dimensions if $n 
\neq 1$ or $m \neq 1$. The spectrum between the $(1,1)$ branes 
and an FZZT brane of parameter $s$ contains a single primary 
field with conformal dimension $h = s^2/4$, in perfect agreement 
with the construction from unitary minimal models. In fact, 
the construction of the $c=1$ brane we have proposed along 
with Cardy's rule implies that there appears a single primary 
with Kac label given by eq.\ (\ref{prescript}) in the 
corresponding spectrum of the $m^{th}$ minimal model. When 
we send $m$ to infinity, this field approaches a $c=1$ primary 
field with momentum 
$$ 2p \ \sim \ [f_sm] + [s] - [f_s m] + \frac{1}{2(m+1)} \, 
          (2[f_s m] + [s]+2) + \mathcal{O}(1/m^2) \ \stackrel{m \rightarrow 
           \infty}{\longrightarrow} \ [s] + f_s \ = \ s \ \ . 
$$ 
Hence, there is perfect agreement between the annulus amplitudes 
in the $c=1$ limit of unitary minimal models and of Liouville 
theory. 
\smallskip 

Let us finally return to the analysis of our FZZT branes at the
Dirichlet points. Since we already understand the effect of shifts
in $s$, we can restrict the following discussion to the case $s=0$.
We would like to probe this FZZT brane with the ZZ brane $(1,1)$. 
The amplitude (\ref{ZZFZZT}) then becomes   
$$ Z_{(1,1),0}(q) \ = \ \eta^{-1}(q) \ = \ 
\lim_{r\rightarrow \infty} Z_{(1,1),(r,r)}(q) \ \ . 
$$ 
In the second equality, we inserted formula (\ref{ann1l}). This 
result nicely confirms our interpretation of the FZZT brane with 
label $s=0$ as an infinite array of point-like branes.

\section{The boundary 2-point function}
\setcounter{equation}{0} 

In this section we shall discuss how the boundary 2-point function
of FZZT branes in Liouville theory can be continued to $c=1$. This
will allow us to derive the exact boundary spectrum, in particular 
the position and width of the expected band gaps, from Liouville 
theory. Our results will confirm nicely the outcome of our previous
discussion in the context of unitary minimal models.  
\medskip   

Let us recall that the boundary 2-point functions for the 
FZZT branes in $c\geq 25$ Liouville theory are given by 
\begin{equation} \label{2ptf} 
\langle \Psi^{s_1 s_2}_{\b_1} (u_1) \Psi^{s_2s_1}_{\b_2} (u_2)  
\rangle \ = \ |u_1 - u_2|^{-2h_{\beta_1}} \, \left[
\, \delta(Q_{b}- \b_1 - \b_2) + \R_{b}^{s_1s_2}(\b_1)\, 
              \delta(\b_1-\b_2) \right] \ \ . 
\end{equation}
The coefficient in front of the first term has been set to $1$ by
an appropriate normalization of the boundary fields. Once this 
freedom of normalization has been fixed, the coefficient of the 
second term is entirely determined by the physics. It takes the 
form 
\begin{equation} \label{reflb} 
\R^{s_1s_2}_b(\b)\  = \  
      \frac{( \pi \mu \gamma (b^2) b^{2-2b^2})^{(Q_{b}-2\b)/2b}}
   {S(\b+ i\ts|b)S(\b-i\ts|b)S(\b + i\ts_1|b)S(\b+i\ts_2|b)}    
   \, \frac{\Gamma_2(2\beta -Q_{b}|b)}{\Gamma_2(Q_{b}-2\beta|b)} 
\end{equation} 
where $\ts = (s_1+s_2)/2$ and $\ts_i = \ts - s_i$. The special 
function $S$ is constructed as a ratio of two Barnes double 
Gamma functions (see appendix A.2 for details). $\R$ is known as 
the {\em reflection amplitude} since it describes the phase shift 
of wave functions that occurs when open strings spanning between 
the FZZT branes with parameters $s_1$ and $s_2$ are reflected by 
the Liouville potential. 
\smallskip 

Our aim now is to continue these expressions to the $c=1$ model (see 
also \cite{Gutperle:2003xf} for a first attempt in this direction). 
Using once 
more the general formula (\ref{Gasym}) for the behavior of Barnes' 
double Gamma function at $b \sim i$, it is not hard to see that 
the boundary reflection amplitude $\R$ of Liouville theory 
vanishes outside a discrete set of open string momenta. 
This behavior is unacceptable since it does not allow for a 
sensible physical interpretation. Actually, it does not originate 
from the physics of the model but rather from an inappropriate 
choice of the limiting procedure. To see this we recall that the 
boundary spectrum of the theory is expected to develop gaps of 
finite width. For such gaps to emerge from Liouville theory, it 
is necessary that the corresponding states in the Liouville model 
become non-normalizable. But our normalization of boundary fields
was chosen such that all states in the boundary theory have unit 
norm. Hence, the normalization we have chosen above is clearly 
not appropriate for the limit we are about to take.
\smallskip 

There exists a distinguished normalization of the boundary fields
in which the reflection amplitude of the boundary theory becomes 
trivial. Since the usual `Liouville wall' ceases to exist at the 
point $b=i$, the trivialization of the boundary reflection amplitude 
seems quite well adapted to our physical setup. Explicitly, the new 
normalization of the boundary fields is given by 
\cite{Ponsot:2001ng,Teschner:2003qk}, 
\begin{equation}\label{normalisation}
\tilde \Psi^{s_1s_2}_{\beta}(u)\ =\ g_{\beta}^{s_1s_2} 
\Psi^{s_1s_2}_{\beta}(u)\ \ ,
\end{equation}
\def\Ga{\Gamma}
\def\ga{\gamma}  
with 
\begin{equation}
g_{\b}^{s_1s_2}\ = \ 
\frac{\big(\pi \mu\ga(b^2)b^{2-2b^2}\big)^{\frac{\b}{2b}}}
     {\Ga_2(Q_{b}-\b-i\ts|b)} \frac{\Ga_2(Q_{b}|b)\Ga_2(Q_{b}-2\b|b)
       \Ga_2(Q_{b}+i s_1|b)\Ga_2(Q_{b}-is_2|b)}
      {\Ga_2(Q_{b}-\b+i\ts_2|b)\Ga_2(Q_{b}-\b+i\ts_1|b)
     \Ga_2(Q_{b}-\b+i\ts|b)}\ \ . 
\nonumber \ \  
\end{equation} 
It is straightforward to rewrite the 2-point function in terms of 
the fields $\tilde \Psi^{s_1s_2}_{\beta}(u)$. The result is  
\begin{eqnarray} \label{new2ptf} 
\langle \tilde \Psi^{s_1 s_2}_{\b_1} (u_1) \, 
\tilde \Psi^{s_2s_1}_{\b_2}  (u_2)  
\rangle & \sim & \frac{\left(\pi \mu \ga(b^2)\right)^{(Q_b/2 
- \b_1 -i\tilde s)/b}}{
  |u_1 - u_2|^{2h_{\beta_1}}} \ C_b^{-1} (\b_1,\s_1,\s_2)\, 
\times \\[2mm] 
 & & \hspace*{-4cm} \times\, \frac{\Gamma(1+2ibp_1) \Gamma(2ib^{-1} p_1)}
 {\Gamma(1-ibs_1)\Ga(-ib^{-1}s_1)\Ga(1-ibs_2)\Ga(-ib^{-1}s_2)}  
 \ \left[\,  \delta(Q_{b}- \b_1 - \b_2)  +  \delta(\b_1-\b_2)\, \right] \ \ . 
\nonumber \end{eqnarray} 
Here we use $\b = Q_{b}/2 + i p$ and the term $C_b$ in the denominator
denotes the bulk 3-point couplings of Liouville theory evaluated 
at the points $\b_1$ and $\sigma_i = Q_{b}/2 + i s_i/2$. In writing this
expression we have omitted a factor that is constant in the momentum 
$p_1$ and invariant under the exchange of $s_1$ and $s_2$ since this
can easily be absorbed in a redefinition of the boundary fields. 
\smallskip 

When rewritten in the new normalization, the only nontrivial term in 
the expression (\ref{new2ptf}) is the bulk 3-point function. Hence, we 
can use the results of \cite{Schomerus:2003vv} (see also section 2) to 
continue the boundary 2-point function to $c=1$. Note, however, that 
the bulk 3-point coupling needs to be inverted. This is possible 
whenever it is non-zero. Since the coupling $C_{c=1}(\b,\s_1,\s_2)$ 
vanishes whenever the factor $P(2p,s_1,s_2)$ does, we conclude that 
boundary fields $\Psi^{s_1s_2}_\b$ exist for $p \in {\cal S}_{s_1s_2} 
\subset \QR$ where
$$ \S_{s_1s_2} \ := \ \{ p \in \QR \, | P(2p,s_1,s_2) \neq 0\} 
     \ \ . 
$$ 
Fields which are labeled by momenta outside this range $\S_{s_1s_2}$ 
do not correspond to normalizable states of the model. On a single 
brane with label $s = s_1 = s_2$ we obtain in particular
$$ \S_{ss} \ = \ \S_{s} \ = \{ p \in \QR\, | \, -\min(f_s,1-f_s) 
      < p < \min(f_s,1-f_s)\ \mod \ 1 \, \}\ \ . 
$$ 
Here, $f_s = s-[s]$ denotes the fractional part of the boundary 
parameter as before. This is exactly the same spectrum we found in 
our discussion of boundary conditions in the $c=1$ limit of unitary 
minimal models ! 
\smallskip 
    
Our result for the 2-point functions of boundary fields in the 
spectrum of the model is given by    
\begin{eqnarray} \label{new2ptfa} 
\langle \tilde \Psi^{s_1 s_2}_{p_1} (u_1) 
     \tilde \Psi^{s_2s_1}_{p_2} (u_2)  
\rangle^{c=1}  & \sim &  
  \frac{e^{-Q(iP_1,is_1/2,is_2/2)} }{|u_1 - u_2|^{2 p^2_1}}   
\left|\frac{\sin\pi s_1 \, \sin \pi s_2}{\sin 2\pi p_1}\right|\,  
 \left[ \delta( p_1 + p_2)  +  
   \delta(p_1-p_2)\right] \nonumber \\[2mm] 
\mbox{ where } \ \ e^{Q(\a_1,\a_2,\a_3)}  & := &  
\frac{\Y(1+2i\ta|1)}{\Y(1|1)}  
\ \prod_{j=1}^3 \,  \frac{\Y(1+2i \ta_j)}{\Y(1+2i\a_j)}\ \ .
\nonumber \end{eqnarray}
In a final step, we change the normalization of the fields $\tilde 
\Psi$ in the $c=1$ theory again so that they possess unit norm, 
\begin{equation}\label{newnormalisation}
\Psi^{s_1s_2}_{p}(u)\ :=\ \tilde g_{p}^{s_1s_2} 
\tilde \Psi^{s_1s_2}_{p}(u)\ \ ,
\end{equation}
with 
\begin{equation}
\tilde g_{p}^{s_1s_2}\ = \ 
\frac{\left(\pi \mu_{\text{ren}}\right)^{-p/2}}{\Ga(-2p)} \, 
 \frac{\Ga_2(1|1) \Ga_2(1+2p|1) \Ga_2(1-s_1|1) \Ga_2(1+s_2|1)
 \Ga(s_1) \Ga(1-s_2)}
{\Ga_2(1+p+\tilde s|1) \Ga_2(1+p-\tilde s_1|1) \Ga_2(1+p-\tilde
 s_2|1) \Ga_2(1+p-\tilde s|1)} \ \ .  
\nonumber \end{equation} 
In an abuse of notation, we have denoted the normalized boundary 
fields of the $c=1$ model by the same letter $\Psi$ as in $c\geq 25$ 
Liouville theory. Nevertheless it important to keep in mind that they
are not justs limits of the corresponding fields in ordinary Liouville 
theory. After this final change of normalization, the boundary 2-point 
functions take the form\footnote{Following the discussion on
normalization of the vacuum after eq.\ \eqref{bulk3ptfunc}, disc
correlators should be rescaled by a factor $\frac{1}{\epsilon}$ so
that the limit of the properly renormalized 2-point function is finite.} 
\begin{eqnarray} \label{2ptfctres1} 
& & \langle \Psi^{s_1 s_2}_{p_1} (u_1) \Psi^{s_2s_1}_{p_2}(u_2) 
  \rangle^{c=1}  \ = \  \frac{{\cal N}^{s_1 s_2}(p)}
{|u_1-u_2|^{2p^2_1}} \ \left[ 
 \delta(p_1+p_2) + \R^{s_1s_2}_{c=1}(p_1) \delta(p_1-p_2)
 \right] \\[3mm] 
& & \hspace*{-.5cm} \mbox{ where }\ \  \R^{s_1s_2}_{c=1}(p) \ = \ 
    \R^{is_1\ is_2}_{c=25} (1-p)^{-1} 
\ \  \mbox{and }  \ \ {\cal{N}}^{s_1s_2}(p) \ = \ 
-2\epsilon p_1 (-1)^{[2p_1]+[s_1]+[s_2]}\nonumber     
\end{eqnarray}  
for all $p_1,p_2 \in \S_{s_1s_2}$. Note that we have been able
to express the ``boundary reflection amplitude'' of the $c=1$ 
theory through the corresponding quantity of the $c=25$ model. 
We shall comment on this interesting outcome of our computation 
in the section 6. 

\section{The bulk-boundary coupling constants}
\setcounter{equation}{0} 

Our final aim is to compute the bulk-boundary couplings of the 
$c=1$ boundary Liouville models. For Liouville theory with real 
$b$, expressions for these couplings were provided by Hosomichi
in \cite{Hosomichi:2001xc}. We shall depart from these formulas to 
derive the corresponding couplings in the $c=1$ model. The 
analysis turns out to be rather intricate. Nevertheless, the 
final answer is in some sense simpler than for $c\geq 25$.
\medskip 
   
Let us begin by reviewing Hosomichi's formula for the bulk-boundary
correlation function
\[
\langle \Phi_{\alpha} (z,\bar{z})\tilde \Psi^{ss}_{\beta} (u)\rangle 
\ =\ |z-\bar{z}|^{h_{\beta}-2h_{\alpha}} 
|z-u|^{-2h_{\beta}} \tilde B (\alpha,\beta ,s) 
\]
of an extended branes labeled by a parameter $s$. In our 
normalization~(\ref{normalisation}) for the boundary fields
$\tilde \Psi$ the couplings $\tilde B$ read
\begin{multline}
\tilde B (\alpha ,\beta ,s)\ =\ 2\pi (\mu \pi \gamma
(b^{2})b^{2-2b^{2}})^{\frac{1}{2b} (Q_{b}-2\alpha)}\,   I (\alpha
,\beta ,s,b) \\[2mm] 
\times \ \frac{\Gamma_{2}
(Q_{b}-\beta|b)\Gamma_{2} (2Q_{b}-2\alpha -\beta|b)\Gamma_{2} (2\alpha
-\beta|b)\Gamma_{2} (Q_{b}+is|b)\Gamma_{2} (Q_{b}-is|b)}{\Gamma_{2}
(Q_{b}-\beta -is|b)\Gamma_{2} (Q_{b}-\beta +is|b)\Gamma_{2}
(\beta|b)\Gamma_{2} (2\alpha|b)\Gamma_{2} (Q_{b}-2\alpha|b)} \label{bbcorrf}
\end{multline}
where $I (\alpha ,\beta ,s)$ is given by the integral
\begin{equation}
I (\alpha ,\beta ,s,b )\ =\ \int_{-\infty}^{\infty} 
 dt\, e^{2\pi ist}\ \frac{S(\tfrac{\beta +2\alpha -Q_{b}}{2}+it|b)
                       S(\tfrac{\beta +2\alpha -Q_{b}}{2}-it|b)}
                       {S(\tfrac{2\alpha -\beta+Q_{b}}{2}+it|b)
                        S(\tfrac{2\alpha -\beta+Q_{b}}{2}-it|b)} \ \ . 
\end{equation}
While sending $b$ to $b = i$, an infinite number of poles and zeroes 
of the integrand approach the integration path. It turns out to be more 
convenient to evaluate the integral by Cauchy's theorem. For $s>|\Im 
\beta |$ we can close the integration contour and rewrite the sum over 
residues as a combination of basic hypergeometric series (see 
appendix A.1),
\begin{align}
&I (\alpha ,\beta ,s,b)\ = \ {}_{2}\Phi_{1} \big( e^{2\pi  ib (\beta +2\alpha
-Q_{b})},e^{2\pi  ib\beta},e^{4\pi  i b\alpha};e^{2\pi
ib^{2}};e^{-2\pi ib (\beta -Q_{b}-is)}\big) \nonumber \\[2mm]
&\quad  \times \ {}_{2}\Phi_{1} \big( e^{-2\pi  ib^{-1} (\beta +2\alpha
-Q_{b})},e^{-2\pi  ib^{-1}\beta},e^{-4\pi  i b^{-1}\alpha};e^{-2\pi
ib^{-2}};e^{2\pi ib^{-1} (\beta -Q_{b}+is)}\big) \nonumber \\[2mm] 
& \quad \times \ e^{-\pi s (\beta +2\alpha
-Q_{b})}S (\beta|b)S (\beta +2\alpha -Q_{b}|b)S (Q_{b}-2\alpha|b) + \
(\alpha\rightarrow Q_{b}-\alpha)\ \ .\label{integral}
\end{align}
Our previous results allow us to estimate the behavior of the double 
Gamma functions and its close relative $S$ in the vicinity of $b=i$. 
But the function ${}_2\Phi_1(A,B,C;\tilde q;z)$ has not appeared in 
our discussion before. Hence, in order to continue our analysis, we 
need to study the basic hypergeometric series in the limit when the 
parameter $\tilde{q}$ approaches $\tilde q = 1$. We shall begin with 
a separate treatment of the two factors $_{2} \Phi_{1}$ before we 
combine the results to address the full coupling $\tilde B$. 
\smallskip 

In dealing with one of the  basic hypergeometric series, the basic 
idea is to approximate the series (we denote the summation parameter 
by $k$) through an integral and to evaluate 
the latter using a saddle-point analysis. We will not give 
a rigorous derivation but shall content ourselves with a rough 
sketch of the argument. This will help to understand the main 
features of our final formulas. Let us introduce the variable 
$y=\tilde{q}^{k}$. Using the asymptotic behavior of the 
q-Pochhammer symbols (see appendix B) we obtain 
\begin{equation}\label{qhypasym}
_{2}\Phi_{1} (A,B,C;\tilde{q};z) \ \sim \ \frac{1}{\sqrt{-2\pi
\log \tilde{q}}}\int \frac{dy}{y}e^{\frac{1}{\log \tilde{q}}h (y)}f (y) 
\end{equation}
where 
\begin{eqnarray*}
h (y)\ =\ h (A,B,C,z,y) & = & \Li (A)-\Li (Ay)+\Li (B)-\Li (By)
\\[2mm] 
& & -\Li (C)+\Li (Cy)-\Li (1)+\Li (y)+\log y \, \log z \ \ ,
\end{eqnarray*}
and
\begin{equation} 
f(y)\ =\ f(A,B,C,z,y) \ = \ 
  \left(\frac{(1-A) (1-B) (1-Cy)}{(1-Ay) (1-By) (1-C) (1-y)}
\right)^{1/2}\ \ .
\end{equation} 
The asymptotics of the integral in eq.\ (\ref{qhypasym}) can be determined
by the method of steepest descent. The leading contribution comes from
saddle-points $y_{i}$, i.e.\ from points satisfying the condition $h' 
(y_{i})=0$. An elementary computation shows that such saddle-points
are obtained as a solution of the quadratic equation  
\begin{equation}\label{saddlepeq}
z (1-Ay) (1-By)\ = \ (1-Cy) (1-y)\ \ .
\end{equation}
Once we have understood which saddle-points $y_{i}$ contribute 
to the integral, we end up with an expansion of the form 
\[
_{2}\Phi_{1} (A,B,C;\tilde{q},z) \ \sim \ \sum_{i} \frac{f(y_{i})}{y_{i}
   \sqrt{h'' (y_{i}) }} e^{\frac{1}{\log \tilde{q}}h (y_{i})} \ \ .
\]
The most difficult part, however, is to find the right saddle-points. 
To begin with, the saddle-point equation (\ref{saddlepeq}) has two 
different solutions, at least for generic choice of the momenta. But 
this is not the full story  since the function $h$ is a complex 
function with branch-cuts. Hence, the description of the saddle point 
$y_i$ is not complete before we have specified on which branch the 
relevant saddle-point is located. We were not able to solve the 
problem in full generality, but for the combination of parameters 
in the problem at hand, we could find the saddle-points through a 
comparison with numerical studies (Mathematica). 
\smallskip 

Let us state the result for the first q-hypergeometric function
appearing in eq.\ (\ref{integral}). Here, the parameter $\tilde{q}$ 
is $e^{2\pi ib^{2}}$ and thus the prefactor $1/\log \tilde{q}$ 
of $h$ in the exponent becomes $-\frac{1}{\epsilon }-\frac{i}{2\pi}$, where
$\epsilon =2\pi iQ_{b}/b$ as introduced in section 2. 
Note that the arguments $A,B,C$ and $z$ depend on $b$ and thus on 
$\epsilon$ and hence they need to be expanded 
around $\epsilon =0$. We shall use the symbols $A_0,B_0,C_0$ and 
$z_0$ to denote the leading terms, i.e. 
\[
A_0 \ = \ B_0 C_0\  \ ,\ \ B_0\ = \ e^{-2\pi  ip_{\beta }}\ \ ,\ \  
C_0\ = \ e^{-4\pi  i p_{\alpha}}\ \ ,\ \  z_0\ =\ e^{-2\pi i (s-p_{\beta })}
\ \ ,\]
where we have set $\alpha=Q_{b}/2+ip_{\alpha}$ and $\beta
=Q_{b}/2+ip_{\beta }$. Since the function $h$ in the exponential 
comes with an extra factor $1/\epsilon$, the sub-leading term in 
the $\epsilon$-expansion of $h$ contributes an extra factor which 
we shall combine with $f$ into a new function $g$. Before we spell 
out our results on the saddle points, we also have to specify the 
branch we use for $\log z$. We found it most convenient to choose 
$\log z \to 2\pi i (p_{\beta }-[p_{\beta}]-s+[s])$. In the end we 
obtain, 
\[  
 {}_{2}\Phi_{1} \big( e^{2\pi  ib (\beta +2\alpha
-Q_{b})},e^{2\pi  ib\beta},e^{4\pi  i b\alpha};e^{2\pi
ib^{2}};e^{-2\pi ib (\beta -Q_{b}-is)}\big)
 \ \sim \ \sum_{i=\pm} \frac{g (y_{i})}{y_{i}\sqrt{h''
(y_{i})}} e^{\big(-\frac{1}{\epsilon}-\frac{i}{2\pi }\big) h (y_{i})}
\]
with
\[
g (y)\ =\ \left(\frac{1-A_0}{1-A_0y}
\right)^{-p_{\alpha}-p_{\beta}/2}
\left(\frac{1-B_0}{1-B_0y} \right)^{-p_{\beta}/2}
\left(\frac{1-C_0}{1-C_0y} \right)^{p_{\alpha}+1/2}
(1-y)^{-1/2}y^{\frac{1+s-p_{\beta}}{2}} \ \ .
\]
The two solutions of the saddle-point equation are denoted by $y_{\pm
}$, and they are given explicitly by  
\begin{align*}
&y_{\pm}\ =\ y_{\pm }(p_{\alpha},p_{\beta},s)\\[2mm]
&\ \ =\ \left\{\begin{array}{l}
e^{\pi  i (2p_{\alpha }+p_{\beta })}\frac{\cos (2\pi
p_{\alpha})\sin (\pi s)}{\sin (\pi (p_{\beta}+s))}\bigg( 1\mp i\tan (2\pi
p_{\alpha})\sqrt{1-\frac{\sin (\pi p_{\beta})^{2}}{\sin (\pi s)^{2}\sin
(2\pi p_{\alpha})^{2}}} \bigg)\\[1mm]
\quad \quad   \text{for}\ |\sin (\pi p_{\beta})|<|\sin (\pi
s)\sin (2\pi p_{\alpha})|\\[3mm]
e^{\pi  i (2p_{\alpha }+p_{\beta })}\frac{\cos (2\pi
p_{\alpha})\sin (\pi s)}{\sin (\pi (p_{\beta}+s))}\bigg(1\mp 
\sqrt{\frac{\sin^{2} (\pi p_{\beta})-\sin^{2} (2\pi
p_{\alpha})\sin^{2} (\pi s)}{\sin^{2}(\pi s)\cos^{2} (2\pi p_{\alpha})}}
\bigg)\\[1mm]
\quad \quad   \text{for}\ |\sin (\pi s)|\geq  |\sin (\pi p_{\beta})|\geq |\sin (\pi
s)\sin (2\pi p_{\alpha})| \\[3mm]
e^{\pi  i (2p_{\alpha }+p_{\beta })}\frac{\cos (2\pi
p_{\alpha})\sin (\pi s)}{\sin (\pi (p_{\beta}+s))}\bigg(1\mp \frac{\sin
(\pi (s+2p_{\alpha}))}{\sin (\pi s)\cos (2\pi p_{\alpha})}
\sqrt{\frac{\sin^{2} (\pi p_{\beta})-\sin^{2} (2\pi
p_{\alpha})\sin^{2} (\pi s)}{\sin ^{2} (\pi (s+2p_{a}))}}
\bigg)\\[1mm]
\quad \quad   \text{for}\ |\sin (\pi p_{\beta})|\geq |\sin (\pi
s)| 
\end{array} \right.
\end{align*}
The last of these three different cases concerns the band gaps, i.e.\ 
momenta $p_\beta \not \in \S_s$. The first two cases, on the other 
hand, apply to the bands so that all bands appear to be split into 
an inner (first case) and an outer (second case) region, depending on 
the value of the bulk momentum $p_\a$. Analyzing the limit in the outer 
part of the bands turns out to be the most difficult task because 
the two saddle-points $y_\pm$ are of the same order, i.e.\ $\Re h 
(y_{+})=\Re h (y_{-})$, and there are at least some subsets in 
momentum space where they both contribute. Within the inner region 
and the gap, the saddle-point $y_+$ dominates so that only the summand 
$i=+$ remains in our asymptotic expansion. The exact description of 
our findings for the intermediate region is rather cumbersome. 
Fortunately, the distinction between inner and outer parts of 
the band disappears once we combine our two basic hypergeometric 
series into the bulk-boundary coupling. Since our main goal is 
to provide a formula for the $c=1$ limit of $\tilde B$, we shall
not attempt to present our findings for $_{2}\Phi_1$ in the outer 
region. Instead, let us recall from above that we also need to 
specify the branch  on which the saddle-points sit. The prescription 
is as follows: If $|y_{i }|>1$, then we choose the branch by moving it 
from the fundamental branch first to $|y_{i }|\pm i0_{+}$,  
infinitesimally above/below the real axis depending on 
whether the fractional part $s-[s]$ is greater or smaller than $1/2$. Then 
we move it on the shortest arc to $y_{i}$.
\smallskip

Having completed the analysis for the first hypergeometric function 
we now turn to the second for which an analogous investigation gives 
the following results,  
\[
\tilde h (y)\ =\ h (A_0,B_0,C_0,z_{0}^{-1}B_0^{-2},y)
\]
and 
\[
\tilde g (y)\ =\ \left(\frac{1-A_0}{1-A_0y}
\right)^{p_{\alpha}+p_{\beta}/2}
\left(\frac{1-B_0}{1-B_0y} \right)^{p_{\beta}/2}
\left(\frac{1-C_0}{1-C_0y} \right)^{-p_{\alpha}+1/2}
(1-y)^{-1/2}y^{\frac{1+s+p_{\beta}}{2}} \ \ .
\]
Note that in this case $\frac{1}{\log \tilde{q}}$ is directly given 
by $-\frac{1}{\epsilon }$. The relevant saddle-points are 
\[
\tilde y_{\pm} \ =\ y_{\pm } (p_{\alpha},p_{\beta},-s) \ \ . 
\]
Our discussion of the three different cases and the choice of 
branches carries over from the previous discussion without any 
significant changes. 

Before we return to the study of $\tilde B$ it is necessary to make one
more  important observation which relates the two saddle-points $y_{\pm}$ 
and $\tilde y_{\pm}$. It is not difficult to see that 
$(A_{0}\tilde y_{\pm})^{-1}$ solves the saddle-point equation
(\ref{saddlepeq}) of $y_{\pm}$. Indeed we find for $p_{\beta}\in
\S_{s}$ that $y_{\pm} = (A_{0} \tilde{y}_{\mp })^{-1}$. 
This fact along with some simple formulas in appendix A.3 allows to 
re-express the final formulas entirely in terms of $y_+(p_\a,p_\b,s)$. 
\medskip 

Now we have all ingredients at our disposal to analyze the asymptotic
behavior of the bulk-boundary correlator. We must combine the divergences 
coming from the q-hypergeometric functions with the divergences of the 
prefactors in eq.\ (\ref{bbcorrf}). Since we know already that boundary 
fields with $p_\b \not \in \S_s$ do not exist, we can restrict our
evaluation of $\tilde B$ to the inner and outer band regions (see 
discussion above).   

In the inner region, only one saddle-point contributes and the divergent 
terms may be determined easily from the formulas we have provided. It 
then turns out that they cancel each other thanks to a rather non-trivial 
dilogarithm identity which is attributed to Ray \cite{Ray:1991}. For
convenience we state Ray's identity in appendix A.3 (eq.\
(\ref{5term})). Hence, we are left with a finite result. After a
tedious but elementary computation we obtain 
\begin{align*}
&\lim_{b\to i }\tilde{B} (\alpha ,\beta ,s)\ =\ 4\pi i (\pi
\mu_{\text{ren}})^{-p_{\alpha}}e^{-\pi i s (p_{\beta}+1)}
\, \sin \pi s \ \bigg(\frac{e^{-2\pi i p_{\alpha} (s+1)}}{1-e^{-4\pi
ip_{\alpha }}} \times \\
&\ \times  \ \frac{g
(y_{+})^{2}e^{-\frac{ih (y_{+})}{2\pi}}}{y_{+}^{2}h'' (y_{+})}\ \frac{\Gamma_2 
(1+p_{\beta}+s|1)\Gamma_2 (1+p_{\beta}-s|1)\Gamma_2
(1-2p_{\alpha}|1)^{2}}{\Gamma_2 (1-2p_{\alpha}-p_{\beta}|1)\Gamma_2
(1-2p_{\alpha}+p_{\beta}|1)\Gamma_2 (1-s|1)\Gamma_2 (1+s|1)} \\
&\quad \quad \quad \ -\
(p_{\alpha}\to -p_{\alpha})\bigg)\ \  
\end{align*}
where $y_+ = y_+(p_\a,p_\b,s)$.  In the derivation we used the fact that 
$y_{+}$  and $(A_{0}\tilde{y}_{+})^{-1}=y_{-}$ are the two solutions of the 
same quadratic saddle-point equation along with a few elementary 
identities which we list in appendix A.3. 

Within the outer regions of the bands, the q-hyper\-geometric functions 
can have contributions from both saddle-points. It is possible 
to show that out of the four possible terms in our product of two  
q-hypergeometric functions, only two appear at any point in momentum space. 
Moreover, after the divergences of the prefactors have been taken into 
account, only one of the two terms is free of divergences. The 
other keeps a rapidly oscillating factor. The cancellation is again 
due to Ray's identity and the finite contributions come from the 
combination $(y_{+ },\tilde{y}_{+})$. In the end, the result for the 
outer region of the band is therefore given by the same formula as 
for the inner region.

It now remains to change the normalization of our boundary fields. We
pass from the fields $\tilde \Psi$ to $\Psi$ using the prescription
(\ref{newnormalisation}), as before. Our result then reads
\begin{align*}
&B (p_{\alpha } ,p_{\beta } ,s)\ =\  4\pi^{2} i (\pi
\mu_{\text{ren}})^{-p_{\alpha}-p_{\beta}/2}e^{-\pi is 
(p_{\beta}+1)}
\, \Ga^{-1}(-2p_\b) \\[2mm] 
&\quad \times \ \Big(\frac{e^{-2\pi i p_{\alpha} (s+1)}}{1-e^{-4\pi
ip_{\alpha }}} \frac{g
(y_{+})^{2}e^{-\frac{ih (y_{+})}{2\pi}}}{y_{+}^{2}h'' (y_{+})}\frac{\Gamma_2 
(1|1)\Gamma_2 (1+2p_{\beta}|1)\Gamma_2
(1-2p_{\alpha}|1)^{2}}{\Gamma_2 (1-2p_{\alpha}-p_{\beta}|1)\Gamma_2
(1-2p_{\alpha}+p_{\beta}|1)\Gamma_2 (1+p_{\beta }|1)^{2}} \\
&\quad \quad \quad \ -\
(p_{\alpha}\to -p_{\alpha})\Big)\ \ .
\end{align*}
This is the formula we anticipated in the introduction. For a direct 
comparison one has to employ the integral representation (\ref{BarnesG}) 
of Barnes' double Gamma function.  
\smallskip

Before we conclude this section, we would like to comment briefly on 
the band-gaps. Actually, we can use the analysis of the bulk-boundary 
couplings to confirm our previous statement that boundary fields 
$\Psi^{ss}_{p_\b}$ with $p_b \not \in \S_s$ decouple from the theory. 
To this end, we shall focus on the coefficients $A$ of the bulk
boundary operator product expansion rather than the couplings $B$. The
former are related to latter by multiplication with the boundary
2-point function. Our claim is that the coefficients $A_{b=i}$ vanish
within the band-gaps. As we have seen before, the 2-point function
contributes a factor that diverges when $p_\b \not \in \S_s$. With
the help of our asymptotic expansion formulas in this section it is
possible to show that the couplings $B$ always diverge slower. 
Hence, boundary fields whose momenta lie in the gaps cannot be excited
when a bulk field approaches the boundary.     

\section{Conclusion and open problems} 
\setcounter{equation}{0} 

In this work we have constructed boundary conditions of the 
Euclidean $c=1$ Liouville model. In particular we argued that 
there exists a family of boundary conditions that is parametrized 
by one real parameter $s \in \QR$. We provided explicit expressions
for their boundary states (see eq.\ (\ref{1ptfceq1})), the boundary 
2-point function (\ref{2ptfctres1},\ref{reflb}), and the bulk-boundary 
coupling (\ref{Bcoupl}). The only quantity that we are missing for a 
complete solution of the model is the boundary 3-point coupling. 
In the case of Liouville theory, the 
corresponding formulas have been found by Ponsot and Teschner in 
\cite{Ponsot:2001ng}. We believe that the analysis of their $c=1$ 
limit can proceed along the lines of the studies we have presented 
in section 5, but obviously this program remains to be carried out 
in detail. 
\smallskip
 
A crucial ingredient in our study was the formula (\ref{Gasym}) for the 
asymptotics of Barnes' double Gamma function. In section 2, the latter 
enabled us to provide a new derivation of the bulk couplings in the 
$c=1$ Liouville model. Our approach here is simpler and more
general than the one that was developed in \cite{Schomerus:2003vv}. With 
this technical progress it is now also possible to calculate the bulk 
couplings of non-rational conformal field theories with $c < 1$. In 
contrast to the $c=1$ case, the models with $c < 1$ are certainly 
non-unitary. We will comment on the results and their relation with 
minimal models elsewhere. Possible applications of such developments
include the 2-dimensional cigar background with $0 \leq k \leq 2$  
\cite{Hikida:2004mp} and similar limits of the associated boundary 
theories \cite{Ribault:2003ss}. 
\smallskip 

All this research, however, was mainly motivated by the desire to 
construct an exact conformal field theory model for the homogeneous
condensation of open string tachyons, such as the tachyon on an 
unstable D0 brane in type IIB theory. As we explained in the 
introduction, the underlying world-sheet theory is a Lorentzian 
version of the model we have considered in this work. But since 
the couplings of our theory are not analytic in the momenta (with 
the boundary state being the only exception), this Lorentzian 
theory cannot be obtained by a simple Wick rotation from the 
solution we described here. Instead, it was suggested in  
\cite{Schomerus:2003vv} to perform the Wick rotation before sending 
$b$ to its limiting value $b=i$. Such a prescription makes 
sense because the couplings of Liouville theory are analytic 
in the momenta as long as $b$ is not purely imaginary. It is 
certainly far from obvious that the Wick rotated couplings 
again possess a well-defined $b=i$ limit. Preliminary studies 
of this issue show that our analysis of the boundary 2-point 
function extends to the Lorentzian case. In the time-like model, 
the spectrum of boundary fields is continuous (there are no 
gaps) and their 2-point function is still given by eq.\ 
(\ref{2ptfctres1}) with a reflection amplitude
$$    \R^{s_1s_2}_{c=1}(x) \ = \ 
    \R^{is_1\ is_2}_{c=25} (1+ix)^{-1}  \ \ .      $$   
This result does not agree with the proposal in \cite{Gutperle:2003xf}.
Note, however, that in the Lorentzian case the correlation functions
can depend on the details of how $b$ approaches $b=i$. Such a behavior 
might be related to a choice of vacuum\footnote{One should compare this 
to the choice of self-adjoint extensions of the Hamilton operator in the 
minisuperspace analysis \cite{Fredenhagen:2003ut}}. Let us also 
anticipate that consistency of our expression for the boundary 
2-point coupling with the half-brane boundary states \cite{Larsen:2002wc,
Gutperle:2003xf} can be checked by means of the modular bootstrap. 
A related observation, though with a somewhat problematic domain of 
open string momenta, has also been made in \cite{Karczmarek:2003xm}. 
Concerning the bulk-boundary coupling, our 
investigations are not complete yet. But before having worked out 
the expression in the Lorentzian model, it is worthwhile comparing 
our formula (\ref{Bcoupl}) for the bulk-boundary coupling with a 
corresponding expression in the time-like theory that was suggested 
recently in \cite{Balasubramanian:2004fz} (eq. [4.14] of that paper). 
In fact, the exponential in the second line of our formula is identical 
to a corresponding factor in the work of Balasubramanian et.\ al. It 
will be interesting to determine the other factors through our 
approach. We shall return to this issue in a forthcoming 
publication. 
\bigskip 
\bigskip 

\noindent 
{\bf Acknowledgements:} We would like to thank V.\ Balasubramanian,
P.\ Etingof, J.\ Fr{\"o}h\-lich, M.R.\ Gaberdiel, K.\ Graham, G.\ Moore, 
B.\ Ponsot, I.\ Runkel, J.\ Teschner and G.\ Watts for very helpful 
discussions and some crucial remarks. Part of this research has been 
carried out during a stay of VS with the String Theory group of
Rutgers University and of both authors at the Erwin Schroedinger
Institute for Mathematical Physics in Vienna. We are grateful for
their warm hospitality during these stays.

\begin{appendix}   
\section{Some background on special functions}
\setcounter{equation}{0} 

In this first appendix we collect a few results on the special 
functions which appear in the analysis of the $c=1$ boundary 
Liouville model. We shall start with some q-deformed special 
functions and then introduce Barnes' double Gamma functions 
and certain closely related special functions. Dilogarithms
and some of their properties are finally reviewed in the third 
subsection. 

\subsection{q-Pochhammer symbols and q-hypergeometric functions} 

One of the most basic objects in the theory of q-deformed special 
functions is the finite q-Pochhammer symbol (see e.g.\ 
\cite{Andrews})  
\[
\fqps{a}{\tilde{q}}{k}\ :=\   \prod_{n=0}^{k-1} (1-a\tilde{q}^{n})\ \ .
\]
Its limit for $k\to \infty$ exists for $|\tilde{q}|<1$ and is denoted by 
\begin{equation}\label{defqps}
\qps{a}{\tilde{q}}\ =\ \prod_{n=0}^{\infty } (1-a\tilde{q}^{n})\ \ .
\end{equation}
The tilde is used to avoid confusion with our convention for $q$ in
the rest of the text.
\smallskip 

With the help of the the finite q-Pochhammer symbol we can now 
introduce the basic hypergeometric series $_{2}\Phi_{1}$. This 
q-deformation of the hypergeometric function $_{2} F_{1} $ is 
defined as 
\[
_{2}\Phi_{1} (A,B,C;\tilde{q};z)\ =\ 
   \sum_{k=0}^{\infty}\frac{\fqps{A}{\tilde{q}}{k}
   \, \fqps{B}{\tilde{q}}{k}}{\fqps{C}{\tilde{q}}{k} 
     \, \fqps{\tilde{q}}{\tilde{q}}{k}}z^{k}\ \ . 
\]
The interested reader can find many basic properties of these 
functions in the literature (see e.g.\ \cite{Mathworld:qhyp,Andrews}). All 
the properties we need in the main text are stated and derived 
there.

\subsection{Barnes' double Gamma function and related functions}
Barnes' double $\Gamma$-function $\Gamma_{2} (x|b)$ is defined
for $x\in \mathbb{C}$ and complex $b$ with $\Re b \not= 0$ (see
\cite{Barnes1}), and can be represented by an integral (for $\Re x>0$), 
\begin{equation}\label{BarnesG}
\log \Gamma_{2} (x|b)\ =\ -\int _{0}^{\infty } \frac{{\rm
d}t}{t} \Bigg( \frac{e^{-Q_{b}t/2}-e^{-xt}}{(1-e^{-bt})
(1-e^{-t/b})}+\frac{(Q_{b}/2-x)^{2}}{2}e^{-t}+\frac{Q_{b}/2-x}{t}  \Bigg) \ \ ,
\end{equation}
whenever $\Re x> 0$. Here we have also introduced the symbol 
\[
Q_{b}=b+b^{-1}\ \ .
\]
Throughout the main text, we often use the following special 
combinations of double Gamma functions, 
\begin{eqnarray} 
\Y(x|b) & =&  \Gamma_{2} (x|b)^{-1}\, \Gamma_{2} (Q_{b}-x|b)^{-1} \\[2mm]
S (x|b) & =&  \Gamma_{2} (x|b) \, \Gamma_{2} (Q_{b}-x|b)^{-1}\ \ .
\end{eqnarray} 
For the latter, we would also like to spell out an integral 
representation that easily follows from the formula (\ref{BarnesG})
above,  
\[
\log S (x|b)\ =\ \int_{0}^{\infty }\frac{dt}{t}\bigg(\frac{\sinh
(Q_{b}-2x)t}{2\sinh (bt)\sinh (t/b)}-\frac{Q_{b}/2-x}{t}  \bigg)\ \ .
\]
This representation is valid for  $|\Re x|<\Re Q_{b}$. Let us also 
note that $S$ is unitary in the sense that $S (x|b)S (Q_{b}-x|b) = 1$ 
The function $S$ is also closely related to Ruijsenaars' hyperbolic Gamma 
function $\Gamma_{h}$ (see \cite{Ruijsenaars:1996}),
\[
S (x)\ =\ \Gamma_{h} \big( i (x-Q_{b}/2) \big)\ \ .
\]
Many further properties of double Gamma functions, in particular 
on the position of their poles and shift properties, can be found 
in the literature (see e.g.\ \cite{Jimbo:1996ss}). 
\smallskip  

Here we would like to prove one relation that involves the 
q-Pochhammer symbols we introduced in the previous subsection. 
To this end, we depart from the integral representation~(\ref{BarnesG}).  
Let us denote the integrand in formula (\ref{BarnesG}) by $f (t,x,b)$, 
s.t.\  $\log \Gamma_{2} (x|b)=-\int_{0}^{\infty}dt\, f (t,x,b)$. The 
function $f$ has the property
\begin{equation}\label{frot}
if (it,x,b)\ =\ -f (t,-ib+ix,-ib) + \frac{(\tfrac{Q_{b}}{2}-x)^{2}}{2t}
(e^{-it}-e^{-t}) \ \ .
\end{equation}
We evaluate the integral by closing the contour in the first quadrant, 
\begin{eqnarray} 
& & \log  \Gamma_{2} (x|b) \ = \  -i\int_{0}^{\infty}dt\, f (it,x,b)-2\pi i
\sum_{\text{Poles}\ t_{n}}\!\!\text{Res}_{t=t_{n}}f (t,x,b) 
\label{tempresult}\\[2mm] 
& = & -\log \Gamma_{2} (-ib+ix|-ib) - \int_{0}^{\infty}
\frac{dt}{2t} (\tfrac{Q_{b}}{2}-x)^{2} (e^{-it}-e^{-t})-2\pi i
\sum_{\text{Poles}\ t_{n}}\!\!\text{Res}_{t=t_{n}} f (t,x,b)
\nonumber\end{eqnarray}
where we have used the formula (\ref{frot}) in passing to the second 
line. The integral in the second term is
\begin{equation}\label{hilfsintegral}
\int_{0}^{\infty}\frac{dt}{t} (e^{-it}-e^{-t})\ =\ -i\frac{\pi}{2}\ \ .
\end{equation}
The poles appearing in the sum over residues are at $t_{n}=2\pi in/b$
with $n=1,2,\dots$ Hence we find
\begin{align*}
2\pi i
\sum_{\text{Poles}\ t_{n}}\text{Res}_{t=t_{n}}f (t,x,b)\ &=\ 
\sum_{n=1}^{\infty} \frac{1}{n}\frac{e^{-Q_{b}\pi in/b}-e^{-2\pi
inx/b}}{1-e^{-2\pi in/b^{2}}}\\
&=\ \sum_{n=1}^{\infty}\frac{1}{n}\frac{(-1)^{n}-e^{\pi
in/b^{2}-2\pi inx/b}}{2i\sin \pi n/b^{2}}\ \ .
\end{align*}
Replacing $\sin \pi n/b^{2}$ through
\[
\frac{1}{2i\sin \pi n/b^{2}}\ =\ e^{-\pi
in/b^{2}}\sum_{m=0}^{\infty}e^{-2\pi inm/b^{2}}
\]
and changing the order of summation, we obtain
\begin{align*}
2\pi i
\sum_{\text{Poles}\ t_{n}}\text{Res}_{t=t_{n}}f (t,x,b)\ &=\ 
\ \sum_{m=0}^{\infty}\big(\log (1-e^{-2\pi im/b^{2}-2\pi i x/b})-\log
(1+e^{-2\pi i (m+1/2)/b^{2}}) \big)\\
&=\ \log  \prod_{m=0}^{\infty}\frac{1-q^{2m}e^{-2\pi ix/b}}{1+q^{2m+1}}
\end{align*}
with $q=e^{-i\pi /b^{2}}$. We can now re-express the infinite product 
with the help of q-Pochhammer symbols (see appendix A.1). Inserting 
this result and eq.\ (\ref{hilfsintegral}) into eq.\ (\ref{tempresult}) 
we finally arrive at
\begin{equation} \label{dGqPoch} 
\Gamma_{2} (x|b)\ =\ e^{\frac{i\pi}{4} (Q_{b}/2-x)^{2}}
\frac{\qps{-q}{q^{2}}}{\qps{e^{-2\pi ix/b}}{q^{2}}}\Gamma_{2}
(-ib+ix|-ib)^{-1}\ \ .
\end{equation} 
This formula is the starting point for our evaluation of the 
asymptotic behavior of $\Gamma_{2}(x|b)$ near $b\to i$. We 
shall return to this issue in appendix B. 

\subsection{Dilogarithm}\label{dilog}

There is one more special function that plays an important role in 
our analysis: Euler's dilogarithm (L.\ Euler 1768). It is defined by
\[
\Li (x)\ :=\ \sum_{n=1}^{\infty }\frac{x^{n}}{n^{2}}\ \ ,
\]
for $|x|\leq 1$. At $x=1$, we find $\Li (1)=\frac{\pi^{2}}{6}$. The 
dilogarithm has the integral representation
\[
\Li (x)\ =\ - \int_{0}^{x}\frac{\log (1-t)}{t}dt \ \ .
\]
It can be analytically continued with a branch cut along the real axis
from 1 to $+\infty $. At $x=1$, the dilogarithm is still continuous,
but not differentiable.
\smallskip 

Properties of Dilogarithms are used frequently to derive the 
formulas that appear in the main text. Here we list the most 
relevant properties
\begin{eqnarray} 
\Li (z^{2}) &  = &  2\Li (z)+ 2\Li (-z)\\[2mm] 
\Li (z)+\Li (1-z)& = & \Li (1)-\log (z)\, \log (1-z)\\[2mm] 
\Li (-z)+\Li (-1/z)& =& 2\Li (-1)-\frac{1}{2}\log ^{2} (z)\\[2mm]
\label{Liclosetoone}
\Li (e^{-\epsilon}) & =& \Li (1)- (1-\log \epsilon)\epsilon + o
(\epsilon) \ \ . 
\end{eqnarray}
Many more properties of the dilogarithm can be found in the 
literature (see e.g.\ \cite{Kirillov:1995en}). Of particular relevance 
for our analysis of the bulk-boundary 2-point function is the 
following equality  
\begin{eqnarray} 
\label{5term} 
\sum_{j=1}^{2}\sum_{l=1}^{2}\big(\Li (\eta_{j}x_{l})-\Li
(\eta_{j}/\zeta_{l})-\Li (\zeta_{j}x_{l})+\Li (\zeta_{j}/\zeta_{l}) 
 \big) & & \\[2mm] 
& & \hspace*{-3cm} =\ \Li (z)-\Li (z\tfrac{\eta_{1}\eta_{2}}{\zeta_{1}
\zeta_{2}})+\log z \log (x_{1}x_{2}\zeta_{1}\zeta_{2}) \ \ .\nonumber 
\end{eqnarray} 
Here, $x_{1,2}$ are the two solutions of the quadratic equation
\begin{equation} \label{quad} 
z (1-\eta_{1}x) (1-\eta_{2}x)\ =\ (1-\zeta_{1}x) (1- \zeta_{2}x)\ \ .
\end{equation} 
Equation (\ref{5term}) is derived from the usual 5-term relation of 
the dilogarithm (see e.g.\ \cite{Kirillov:1995en}) with the help of the 
following list of relations that hold for any two solutions $x_1,
x_2$ of the equation (\ref{quad}), 
\begin{eqnarray}
\left(1-z\, \frac{\eta_1\eta_2}{\zeta_1\zeta_2}\right) (1-\eta_i x_1) 
   (1-\eta_i x_2) & = & \left(1-\frac{\eta_i}{\zeta_1}\right)
                        \left(1-\frac{\eta_i}{\zeta_2}\right)
       \label{eq1} \\[2mm] 
 \left(1-z\, \frac{\eta_1\eta_2}{\zeta_1\zeta_2}\right) (1-\zeta_i x_1) 
   (1-\zeta_i x_2) & = & \frac{(\zeta_i - 
\eta_1)(\zeta_i-\eta_2)}{\zeta_i}
      \label{eq2} \\[2mm] 
  x_1 x_2 & = & \frac{1-z}{\zeta_1\zeta_2 - z \eta_1 \eta_2} \label{eq3}
\ \ .  
\end{eqnarray}   
We leave the details to the reader. Let us note that the properties
(\ref{eq1}) to (\ref{eq3}) are also used frequently to simplify the 
final formula for the bulk-boundary 2-point function.  

\section{Asymptotic behavior of q-Pochhammer symbols} 
\setcounter{equation}{0} 

In order to evaluate the behavior of the double Gamma function 
near $b=i$ one can start from our equation (\ref{dGqPoch}). Since 
the double Gamma function that appears on the right hand side of 
this equation is analytic at $b=i$, the main issue is to understand
the asymptotic behavior of the q-Pochhammer symbol. This is what we 
are concerned with here. To begin with, we shall parametrize $\tilde{q}$ 
by $\log \tilde{q}=- \epsilon$. In general, the limit can depend on the
way we send $\epsilon $ to zero. We start by taking the logarithm of the
definition~\eqref{defqps},
\begin{align*}
\log \qps{a\tilde{q}^{z/2}}{\tilde{q}}\ &= \ \sum_{n=0}^{\infty }\log
(1-a\, \tilde{q}^{z/2}\tilde{q}^{n})\\
&=\ \frac{1}{2}\log (1-a\, \tilde{q}^{z/2})+\int_{0}^{\infty }\log
(1-a\, \tilde{q}^{z/2+x})dx\\
&\quad +\int_{0}^{\infty }\phi_{1} (t-[t])\frac{d}{dx}\big(\log
(1-a\,\tilde{q}^{z/2+x}) \big)\big|_{x=t}\, dt \ \ .
\end{align*}
Here, we used the Euler-MacLaurin sum formula to re-express the sum as
an integral, $\phi_{1}$ is the first Bernoulli polynomial, 
\[
\phi _{1 } (t)\ =\ \frac{1}{2}-t \ \ .
\]
Note that the branch of the logarithm $\log
(1-a\,\tilde{q}^{\frac{z}{2}+x})$ is chosen such that it approaches
$0$ when we send $x$ along the real axis toward $+\infty $.
The first integral is given by the dilogarithm (see
appendix~\ref{dilog} for the definition and properties), so we obtain
\begin{equation}\label{qpsasymp}
\qps{a\tilde{q}^{z/2}}{\tilde{q}}\ =\ e^{-\frac{1}{\epsilon }\Li
(a\tilde{q}^{z/2})} \   (1-a \tilde{q}^{z/2})^{\frac{1}{2}} \ \exp I
(\epsilon,a,z) \ \ ,
\end{equation}
where $I (\epsilon,a,z )$ denotes the integral containing the
Bernoulli polynomial,
\[
I (\epsilon,a,z)\ :=\ \int_{0}^{\infty }\phi _{1} (t-[t]) \frac{(-a)\log (\tilde{q})\, 
\tilde{q}^{z/2+t}}{1-a\, \tilde{q}^{z/2+t}}\, dt \ \ .
\]
As long as $|a|<1$, $I (\epsilon,a,z)$ is suppressed by $\log
\tilde{q}=-\epsilon $. On the other hand, if $a=1$, then the integral
gives a non-trivial contribution.
Let us explain that in more detail. 
We rewrite the integral as
\[
I (\epsilon,a,z)\ =\ \int_{0}^{\infty } \phi_{1} (\tfrac{t}{\epsilon}-[\tfrac{t}{\epsilon} ])
\frac{1}{a^{-1}\,  e^{\epsilon z/2}\, e^{t}-1}dt \ \ .
\]
When $\epsilon$ goes to zero, the strongly oscillating sign of the 
Bernoulli polynomial suppresses the integral, so that 
generically $I (\epsilon,a,z)$ vanishes in that limit. 
The leading contribution comes from the pole
of the integrand, and if it approaches the path of integration when
sending $\epsilon$ to zero, we can get a finite contribution. For
$|a|<1$, the pole is always outside of the integration region, but for
$|a|\geq 1$, the pole might come close to the real axis. We can take
this effect into account by extracting the pole part from the integrand,
\[
\frac{1}{a^{-1}\,  e^{\epsilon z/2}\, e^{t}-1} \ =\ \frac{1}{t-\log a
+\epsilon z/2} + \text{regular}\ \ .  
\]
The regular part will not contribute to the integral 
in the limit $\epsilon\to 0$, so we can rewrite $I (\epsilon,a,z)$ as
\[
I (\epsilon,a,z)\ =\ \int_{0}^{\infty }\phi_{1}
(t-[t])\frac{1}{t-\frac{\log a}{\epsilon}+\frac{z}{2}} + o (\epsilon^{0})\ \ .
\]
The remaining integral is given by the difference of $-\log \Gamma
(-\frac{\log a}{\epsilon}+\frac{z}{2})$ and the first terms of 
its Stirling series,
\begin{multline*}
I (\epsilon,a,z)\ =\ -\log \Gamma (-\tfrac{\log
a}{\epsilon}+\tfrac{z}{2})+\tfrac{1}{2}\log 2\pi\\
+\big(-\tfrac{\log a}{\epsilon}+\tfrac{z}{2}
\big)\Big(\log \big( -\tfrac{\log a}{\epsilon}+\tfrac{z}{2}\big)-1
\Big) -\tfrac{1}{2}\log \big(-\tfrac{\log a}{\epsilon}+\tfrac{z}{2}
\big) + o (\epsilon^{0})\ \ .
\end{multline*}
As long as $|a|\leq 1$ and $a\not= 1$, the limit of $I$ vanishes because the
$\Gamma$-function approaches  its Sterling approximation. If 
$a=1$, we find 
\[
\exp I (\epsilon,1,z) \ =\ \frac{\sqrt{2\pi}}{\Gamma (z/2)}\bigg(
\frac{z}{2}\bigg)  ^{\frac{z-1}{2}} e^{-\frac{z}{2}}\   \big(1+o
(\epsilon^{0})\big) \quad \text{for}\  z\in \mathbb{C}\setminus (-\infty ,0] \ \ . 
\]
This formula together with~\eqref{qpsasymp} describe the asymptotics
of q-Pochhammer symbols in the cases we shall need
for our applications.
\medskip

In the following we want to apply the gained insight in a number of
special cases. First, let us write down the full
asymptotics~\eqref{qpsasymp} for $a=1$,
\[
\qps{\tilde{q}^{z/2}}{\tilde{q}}\ =\ e^{-\frac{1}{\epsilon}\Li
(1)} \epsilon^{-(z-1)/2} \frac{\sqrt{2\pi}}{\Gamma (z/2)}
(1 +o (\epsilon^{0}))\ \ 
\]
where we used~\eqref{Liclosetoone} from appendix~\ref{dilog} 
to expand the dilogarithm close to
1. We observe that this expression has zeroes for $z=-2,-4,-6,\dots
$ which is consistent with the general definition~\eqref{defqps}.
For integer $z$ these asymptotics can alternatively be derived using
modular properties of Dedekind's $\eta$-function.
\smallskip

We are often interested in q-Pochhammer-symbols of the form $\qps{f(\tilde{q})}{\tilde{q}}$ with some function $f (\tilde{q})$. In most
cases, it is enough to expand $f$ to first order around $\tilde{q}=1$,
and we find 
\[
\qps{f (\tilde{q})}{\tilde{q}}\ =\ \qps{f (1)\tilde{q}^{f' (1)/f
(1)}}{\tilde{q}} (1+o (\epsilon^{0})) \ \ .
\]
This relation breaks down if $f (1)=1$ and $f' (1)/f (1)=-n$
for some $n=0,1,2,\dots $. 
\smallskip

Let us look at an example. Set $\tilde{q}=e^{-2\pi i/b^{2}}$
and $f (\tilde{q})= e^{-2\pi p/b
(\tilde{q})}\tilde{q}^{z/2}$ for $p\in \mathbb{R}$.
Then we obtain
\begin{equation}
\log \qps{e^{-2\pi p/b}\tilde{q}^{z/2}}{\tilde{q}}\ =\
-\frac{1}{\epsilon} \text{Li}_{2} (e^{2\pi i p}) +\frac{1}{2} 
(1-z-p ) \log (1-e^{2\pi i p })+ o (\epsilon^{0}) \
\ .
\end{equation}
By carefully choosing the correct branch of the logarithm we can
rewrite the result as
\begin{equation}\label{pochasymp}
\qps{e^{-2\pi p/b}\tilde{q}^{z/2}}{\tilde{q}}\ =\
e^{-\frac{1}{\epsilon}\Li (e^{2\pi i p})} \big( 2|\sin \pi p| e^{\pi i
(p-[p]-1/2)}  \big)^{(1-z-p)/2} \  \big( 1+o (\epsilon^{0}) \big) \ \ ,
\end{equation}
where $[p]$ denotes the largest integer smaller or equal $p$. 
\smallskip

Let us finally look at the behavior of $\qps{a}{\tilde{q}}$ when $a$ is
close to 1. Obviously $\qps{1}{\tilde{q}}=0$, and the derivative
is easily obtained as
\begin{equation}\label{atoone}
\frac{d}{da} \qps{a}{\tilde{q}}\Big|_{a=1}\ =\
-\qps{\tilde{q}}{\tilde{q}} \ \ .
\end{equation}
Similar formulas for the asymptotics of q-Pochhammer symbols have been 
derived before (see in particular \cite{McIntosh:1999}).   
\end{appendix}

\end{document}